\documentclass[prd,twocolumn,showpacs,showkeys,preprintnumbers,floatfix,nofootinbib,superscriptaddress]{revtex4-1}
\usepackage{amsfonts} % AMS
\usepackage{amssymb} % AMS
\usepackage{amsmath} % AMS
\usepackage{graphicx} % Include figure files
\usepackage{subfigure} % Include figure files
\usepackage{array} % array
\usepackage{dcolumn} % Align table columns on decimal point
\usepackage{bm} % bold math
\usepackage{latexsym} % latex symbols
\usepackage{longtable} % long tables
\usepackage{hyperref} % hypertext links 
\usepackage{bbold}
\usepackage{color}
\usepackage{multirow}
\usepackage{rotating}
\usepackage{comment}
\usepackage{epstopdf}
\usepackage{epsfig}

\usepackage[dvipsnames,usenames]{xcolor}
\usepackage{tkz-berge,tikz}
\usepackage[qm,braket]{qcircuit}

\definecolor{green}{rgb}{0.1, 0.8, 0.1}

\begin{document}

%%%

\title{Quantum Computing for Neutrino-nucleus Scattering}
\author{Alessandro~Roggero}
\email{roggero@uw.edu}
\affiliation{Institute for Nuclear Theory, University of Washington, 
Seattle, WA 98195, USA}

\author{Andy~C.~Y.~Li}
\email{cli@fnal.gov}
\affiliation{Fermi National Accelerator Laboratory, Batavia, IL 60510}

\author{Joseph~Carlson}
\email{carlson@lanl.gov}
\affiliation{Los Alamos National Laboratory, Theoretical Division T-2, Los Alamos, NM 87545, USA}

\author{Rajan~Gupta}
\email{rajan@lanl.gov}
\affiliation{Los Alamos National Laboratory, Theoretical Division T-2, Los Alamos, NM 87545, USA}

%\author{Alexandru~Macridin}
%\email{macridin@fnal.gov}
%\affiliation{Fermi National Accelerator Laboratory, Batavia, IL 60510}

\author{Gabriel~N.~Perdue}
\email{perdue@fnal.gov}
\affiliation{Fermi National Accelerator Laboratory, Batavia, IL 60510}

%\collaboration{}
%
%
%
\preprint{LA-UR-19-31323}
\preprint{INT-PUB-19-052}
\preprint{FERMILAB-PUB-19-547-QIS}
\pacs{
}
\date{\today}
\begin{abstract}
Neutrino-nucleus cross section uncertainties are expected
to be a dominant systematic in future accelerator neutrino experiments. 
The cross sections are
determined by the linear response of the nucleus to the weak field
of the neutrino, and are dominated by energy and distance scales
of the order of the separation between nucleons in the nucleus.
These response functions are potentially an important
early physics application of quantum computers.
Here we present an analysis of required resources and expected scaling for scattering cross section calculations.
We also examine simple small-scale neutrino-nucleus models on modern quantum hardware. In this paper we use variational methods to obtain the ground state and then implement the relevant time evolution. In order to tame the errors in present-day NISQ devices we explore the use of different error-mitigation techniques to increase the fidelity of the calculations.
\end{abstract}
\maketitle
%
%
%
%\nopagebreak
%
%%%%%%%%%%%%%%%%%%%%%%%%%%%%%%%%%%%%%%%%%%%%%%%%%%%%%%%%%%%%%%%%%%%%%
%%%  SECTION                                                      %%%
%%%%%%%%%%%%%%%%%%%%%%%%%%%%%%%%%%%%%%%%%%%%%%%%%%%%%%%%%%%%%%%%%%%%%
\section{Introduction}
\label{sec:intro}

Establishing the existence of ``CP-violation'' in the lepton sector through neutrino oscillation experiments, and testing the three-flavor neutrino framework at a long baseline experiment such as DUNE \cite{refdune}, are challenging tasks. 
Successful execution of these goals requires very fine controls on systematic uncertainties.
Interaction model uncertainties will likely be the dominant systematic uncertainties in mature experiments, and further theory work is required to bring them under control~\cite{Benhar2017,Alvarez2018}.

Experiments use event generators, such as GENIE~\cite{Andreopoulos:2009rq} NEUT~\cite{Hayato:2002sd}, NuWro~\cite{PhysRevC.86.015505,Juszczak:2005zs,Golan:2012rfa}, and GiBUU~\cite{Buss:2011mx}, to connect  final states observed in the detectors to the detailed underlying kinematics.
There are two defining features for neutrino-nucleus interaction signals.
First, the kinematic details of beam neutrinos are unknown on an event by event basis, and even the overall flux may be poorly constrained.
Second, neutrino experiments favor heavy nuclear target materials to drive up event rates at the price of introducing very complex nuclear physics in the event reactions.

Because they are tools for understanding detector efficiency and backgrounds, event generators must simulate all types of constituents possible in the final state of an interaction and their momenta on an event-by-event basis.
An ideal theory input would provide fully-differential neutrino-nucleus cross sections with respect to the kinematics of every final-state particle, for all combinations of neutrino flavor and helicity, and for every nucleus in the target. Unfortunately, even the most sophisticated modern theory typically provides only the kinematics for the final state lepton, and generally only covers a subset of the experimentally accessible phase space \cite{RevModPhys.87.1067}.

On classical computers, inclusive scattering in ab-initio calculations are obtained via imaginary-time (Euclidean) correlation functions~\cite{Lovato2016,Lovato2018} or in factorization schemes\cite{Rocco2019}. These are typically relevant to inclusive scattering only, though some progress has been made towards exclusive processes.~\cite{Pastore2019} Exact treatments, even for the ground state, scale exponentially in the nucleon number due to the Fermion sign problem. Constrained path algorithms are useful for low-lying states, but
scattering has proven to be intractable on classical computers.

Since its first conceptualization~\cite{Feynman1982}, quantum computing has been seen to offer a potentially powerful tool for computing ab-initio the time evolution of strongly correlated quantum systems, like the ground state of nuclei, with controllable errors~\cite{Lloyd1996}. This is mostly due to the ease of incorporating fundamentally quantum effects like entanglement and interference within it's language, something that in general requires an exponential overhead on classical digital computers.

In an earlier publication~\cite{Roggero2018}, some of us proposed a quantum algorithm for digital quantum computers to efficiently estimate properties of (nuclear) final states in scattering events like neutrino-nucleus reactions using a variant of quantum algorithms developed for quantum chemistry applications~\cite{Somma2002,Wecker2015}. 

In this work, we start by carefully assessing in Sec.~\ref{sec:eft_estimates} the quantum resources needed for a minimally realistic description of a scattering process off a nucleus in the linear response regime. In particular, we first provide detailed implementations of quantum circuits simulating the time evolution needed for the algorithm presented in~\cite{Roggero2018} in Sec.~\ref{sec:general_time_evolution} and also explore the use of an alternative approach in Sec.~\ref{sec:new_scheme_qubit}.
While current quantum computing hardware is insufficient to do these calculations in full for relevant nuclei, especially without active error correction, our goal in Sec.~\ref{sec:actual_implementation} is to demonstrate proof of principle calculations that will motivate further research and development in this area.

\section{Lattice Nuclear Model}
\label{sec:eft_estimates}
In this paper, we study systems using pionless effective field theory\cite{Bedaque2002AnnRev,Hammer2019nuclear} on a lattice to explore quantum computing of nuclei and their response. We have chosen 
pionless effective field theory as it is the simplest possible model of nuclei and their interactions that exhibits some very basic properties of atomic nuclei. 
It consists of non-relativistic nucleons interacting with a contact interaction
that reproduces large scattering lengths at low
energies.

At leading order it has nucleon-nucleon contact interactions,
describing the low-energy s-wave interactions
in spin zero isospin one (S=0, T=1) and spin one isospin zero (S=1, T=0) nucleon pairs.
The measured scattering length in S=0, T=1 (e.g. $nn$ scattering)
is approximately -18 fm, almost
a bound state; while in the S=1 T=0 channel there is a weakly bound state, the deuteron, with a binding energy of 2.225 MeV.
For initial studies, these simple pionless interactions are preferable since they can be efficiently implemented in a lattice basis; 
indeed they have many similarities to a 3D Hubbard model with attractive
interactions, but with four species of fermions (neutrons and protons
with spins up and down).

In addition to the two-nucleon interactions, a three-nucleon interaction is required to avoid collapse into deeply bound states~\cite{Bedaque1999a,Bedaque1999b}.
Pionless effective field theory has been shown to approximately reproduce the binding of three and four nucleon systems, and to nearly produce weakly bound nuclei (with respect to break up into four-particle clusters) for $A=8$ and $A=16$~\cite{Contessi_2017,Dawkins2019clustering}, as seen in nature.
More complex interactions including virtual pions are necessary for
more accurate studies of lepton-nucleon interactions, as these provide fits
to NN scattering data up to momenta of several inverse fermi.

The resulting lattice Hamiltonian for the pionless theory is:
\begin{equation}
\begin{split}
H&= 2DtA -t \sum_{f=1}^{N_f}\sum_{\langle i,j\rangle}^M \left[ c^\dagger_{i,f}c_{j,f} + c^\dagger_{i,f}c_{j,f}\right]\\
&+ \frac{1}{2} C_0 \sum_{f\ne f'}^{N_f}\sum_{i=1}^M n_{i,f}n_{i,f'} \\
&+ \frac{D_0}{6} \sum_{f\ne f'\ne f''}^{N_f}\sum_{i=1}^M n_{i,f}n_{i,f'}n_{i,f''}\;,
\end{split}
\label{eq:lattice_ham}
\end{equation}
where $A$ is the number of nucleons, $D$ the space dimension, $N_f$ the number of fermionic species and $M$ the number of lattice sites.
$C_0$ and $D_0$ describe the strengths of the attractive and repulsive two- and three-nucleon interactions, respectively.  Here we assume the S=0, T=1 and S=1, T=0 scattering lengths are the same. If the box size is $L$ and $M=N_L^D$, the kinetic energy parameter is $t=\hbar^2/2ma^2$ with the lattice spacing $a=L/N_L$. For the calculations presented in this section we use the numerical values reported in Tab~\ref{tab:hparams} (obtained from~\cite{Rokash2013}) and corresponding to a lattice spacing of $a=1.4$ fm.

\begin{table}[]
\begin{tabular}{c|c|c}
$t$ [MeV] & $C_0$ [MeV] & $D_0$ [MeV] \\ \hline
10.5794 & -98.2265511 & 127.839693
\end{tabular}
\caption{Hamiltonian parameters, corresponding to a lattice spacing $a=1.4$ fm, taken from~\cite{Rokash2013}.}
\label{tab:hparams}
\end{table}

We can encode the Fock space with $\Omega=N_f\times M$ fermionic modes into $\Omega$ qubits using the Jordan-Wigner~\cite{JWreference} transformation to obtain the mapping
\begin{equation}
n_q\equiv c^\dagger_q c_q=\frac{1-Z_q}{2}
\end{equation}
and
\begin{equation}
\begin{split}
c^\dagger_{q}c_{p}+c^\dagger_{p}c_{q}&=-\frac{1}{2}X_qZ_{q+1}\cdots Z_{p-1}X_p\\
&-\frac{1}{2}Y_qZ_{q+1}\cdots Z_{p-1}Y_p\;.
\end{split}
\label{eq:JW}
\end{equation}
In Eq.~\eqref{eq:JW}, we use $X_q$,$Y_q$ and $Z_q$ to denote the corresponding Pauli matrix acting on qubit $q$ and the dots indicate Pauli $Z$ matrices on the qubits along the chosen normal ordered path connecting the qubit for orbital $q$ with the qubit for orbital $p$ (for more details see eg.~\cite{Somma2002}).
In this work we order the qubit placing next to each other the $N_f$ qubits representing the same lattice site and different spin-isospin quantum number. This choice (equivalent to the mapping used in early works on quantum chemistry like~\cite{Whitfield2011}) is particularly convenient in our case due to the presence of the 3-body interaction which requires to couple triplets of fermions at the same lattice point. This is a different situation to the one encountered in quantum chemistry where a different mapping focusing on the kinetic energy is usually chosen (see eg.~\cite{Welch2014}).

The nuclear Hamiltonian can now be written entirely in terms of Pauli operators. Starting from the kinetic energy component
\begin{equation}
\begin{split}
K&=-t\sum_{f=0}^{N_f} \sum_{\langle i,j\rangle}^M \left[ c^\dagger_{i,f}c_{j,f} + c^\dagger_{i,f}c_{j,f}\right]\\
&=-t\sum_{f=0}^{N_f} \sum_{i=0}^{M-1}\sum_{j\in{NN(i)}} \left[ c^\dagger_{i,f}c_{j,f} + c^\dagger_{i,f}c_{j,f}\right]
\end{split}
\end{equation}
we find explicitly
\begin{equation}
\begin{split}
K&=\frac{t}{2}\sum_{f=0}^{N_f-1} \sum_{i=0}^{M-1}\sum_{j\in{NN(i)}}\bigg(\\
& X_{N_fi+f}Z_{N_fi+f+1}\cdots Z_{N_fj+f-1}X_{N_fj+f}\\
&+Y_{N_fi+f}Z_{N_fi+f+1}\cdots Z_{N_fj+f-1}Y_{N_fj+f}\bigg)
\end{split}
\end{equation}
where in the expressions above $NN(i)$ are the indices of the nearest neighbors of lattice site $i$. 
In turn, the potential can be written as the following diagonal operator
\begin{equation}
\begin{split}
V=&\frac{M}{4} \left(\frac{N_f(N_f-1)}{2}\right) \left( C_0+\frac{N_f-2}{3}D_0\right)\\
&-\frac{N_f-1}{4}\left(C_0+\frac{N_f-2}{2}D_0\right)\sum_{i=0}^{M-1}\sum_{f=0}^{N_f-1}Z_{4i+f}\\
&+\frac{C_0+(N_f-2)D_0}{4}\sum_{i=0}^{M-1}\sum_{f=0}^{N_f-1}\sum_{f'>f}Z_{4i+f}Z_{4i+f'}\\
&-\frac{D_0}{4}\sum_{i=0}^{M-1}\sum_{f=0}^{N_f-1}\sum_{f''>f'>f}Z_{4i+f}Z_{4i+f'}Z_{4i+f''}\;,
\end{split}
\end{equation}
which, for the common case with $N_f=4$, simplifies to
\begin{equation}
\label{eq:pot_energy}
\begin{split}
V=&M\left( \frac{3}{2}C_0+D_0\right)-\frac{3}{4}\left(C_0+D_0\right)\sum_{i=0}^{M-1}\sum_{f=0}^{3}Z_{4i+f}\\
&+\frac{C_0+2 D_0}{4}\sum_{i=0}^{M-1}\sum_{f=0}^{3}\sum_{f'>f}Z_{4i+f}Z_{4i+f'}\\
&-\frac{D_0}{4}\sum_{i=0}^{M-1}\sum_{f=0}^3\sum_{f''>f'>f}Z_{4i+f}Z_{4i+f'}Z_{4i+f''}\;.
\end{split}
\end{equation}
Note that this operator is composed of a sum of 
\begin{equation}
N_V = M N_f\left( 1+\frac{N_f-1}{2}\left(1+\frac{N_f-2}{3}\right)\right)
\end{equation}
mutually commuting operators while the kinetic energy term is composed of a possibly much larger number of mutually non-commuting operators. In this work we will consider two different breakups of the Hamiltonian: one where we separate all the $N_K=4 D M N_f$ terms in the expansion of the kinetic energy from a single potential energy term and one where we simply separate the kinetic and potential energy terms and treat each one exactly (cf. split-operator step in \cite{kivlichan2019improved}).
In the following we will refer to these splitting as $\alpha$ and $\beta$.

As we have anticipated in the introduction, the main observables we are seeking are semi-exclusive cross sections for a neutrino to scatter off a nucleus. A related but easier to compute quantity of interest is the response function
\begin{equation}
S(\omega) = \sum_f \delta (\omega -(E_f - E_0))
	\langle 0 | O^\dagger | \Psi_f \rangle
	\langle \Psi_f | O | 0 \rangle
\end{equation}
which directly measures the inclusive cross section. The operator $O$ in the above expression represents the electro-weak excitation operator of the incoming neutrino, while $\{\ket{\Psi_f}\}$ and $\{E_f\}$ are the eigenstates and eigenvalues of a nuclear hamiltonian like \eqref{eq:lattice_ham}. In~\cite{Roggero2018} we show how, by a slight modification of quantum algorithms developed for the estimation of $S(\omega)$~\cite{Somma2002,Wecker2015}, one can set up a quantum computation to sample efficiently the most important final states of a neutrino-nucleus collision.
The dominant cost in computing the cross section comes from the need to perform time-evolution and we dedicate the next subsection to characterize, for a realistic setup, how large this cost actually is. We finish this section by exploring an alternative approach based on the technique of qubitization~\cite{Low_and_Chuang_2019} which provides an optimal asymptotic cost.

\subsection{Time evolution}
\label{sec:general_time_evolution}
The cost of our original scheme~\cite{Roggero2018} is dominated by the implementation of the time evolution unitary operator generated by this Hamiltonian controlled with an ancilla qubit. In the following we will estimate the computational cost of the algorithm by looking at the number of expensive operations (CNOTs and single qubit rotations) needed to achieve some target accuracy in the inclusive response. We will account for the ancilla control of the time-evolution unitary by considering every rotation to be a controlled one which we implement in a standard way (see Eq.~\eqref{eq:crotz} in Appendix.~\ref{app:exp_pauli}). Note that we can easily extend parallelization even when rotations are controlled by ancillas as explained in~\cite{Hastings2015}.

In this section we will mostly consider product formulas, in particular we will study in detail both linear and quadratic Trotter-Suzuki break-ups, and comment on the possible beneficial use of qubitization at the end of the section. We remind the reader that other techniques have been developed beside these, an important one being for instance the LCU method and it's variants \cite{Childs2012,Berry2015}. Since the implementation of the LCU method comes with a possibly much larger overhead in qubit count (see eg.~\cite{Childs2018} for a detailed study of a single Hamiltonian) we will not explore its use further in this first work.% apart from assessing the cost of implementing one of its primitives (see Sec.~\ref{app:qub_cost}).

\subsubsection{Product Formulae: number of steps}
Product formulae obtained from the Trotter-Suzuki decomposition~\cite{Trotter59,Suzuki91} are essentially small-time approximations of the time-evolution unitary operator $U(t)=exp(-it H)$ with additive error $\delta_{TS}(t)=\mathcal{O}\left(t^\gamma\right)$ for some $\gamma>1$ (eg. $\gamma=2$ for the linear breakup). This implies that in order to perform a simulation lasting a total time $\tau$ with bounded error we will need to divide the total time interval $[0,\tau]$ into $r$ segments and use in each one the approximate evolution operator to obtain
\begin{equation}
\label{eq:approx_time_ev}
\begin{split}
\delta_\tau = &\|e^{-i\tau H} - \widetilde{U}(\tau/r)^r \| \\
&\leq r \|e^{-i(\tau/r) H} - \widetilde{U}(\tau/r) \| = \mathcal{O}\left(r^{1-\gamma}\right)\;,
\end{split}
\end{equation}
with $\widetilde{U}(t)$ the approximate propagator. In general the norm appearing in Eq.~\eqref{eq:approx_time_ev} is the standard operator (or spectral) norm. For our application we are interested in systems with a fixed number of nucleons and both the Hamiltonian and every single term in either the $\alpha$ and $\beta$ splitting commutes with the baryon number operator. We will consider then a {\it physical} norm defined as
\begin{equation}
\left\| O \right\|_{phys}=\sup\left\{\left\|O\rvert\psi\rangle\right\|_2 : \rvert\psi\rangle \text{ A-baryon state} \right\}\;.
\end{equation}
In other words, physical norms only take into account quantum states which respect the symmetry of the Hamiltonian and the initial conditions.
Note that more generally we could restrict the class of physical states using additional symmetries (ie. isospin) resulting in an even tighter norm since $\left\| O \right\|_{phys} \leq \left\| O \right\|$. This definition is very convenient in our case since it allows us to define a reasonable lower bound for the base time $\tau=2\pi/\Delta H$ that we need in the QPE part of the algorithm. In fact it is sufficient to provide a physical upper bound on the maximum spread in energy attainable in an $A$ body system as
\begin{equation}
\label{eq:energy_spread}
\begin{split}
\Delta H &= E_{max} - E_{min}\\
&= \|K\|_{phys} + \|V_2\|_{phys} + \|V_3\|_{phys} + A b_{max}\;,
\end{split}
\end{equation}
where $b_{max}$ is the nuclear binding energy at saturation density and we've used the estimate $\lvert E_{min}\rvert\leq A b_{max}$ for the lowest energy value. An even better bound can be obtained by considering $\|V\|_{phys} = \|V_2 + V_3\|_{phys}$ which is smaller due to the opposite signs in the interaction terms (see Eq.~\eqref{eq:pot_phys_norm} in Appendix~\ref{app:tstep_details}). 

As a simple starting point we now consider the linear order Trotter-Suzuki product formulae of the form
\begin{equation}
\label{eq:lin_trott}
U^\alpha_L(\tau)=\prod_k^{N_K}e^{-i\tau K_k}e^{-i\tau V}\;,
\end{equation}
for the $\alpha$ splitting, where we used the expansion $K=\sum_k^{N_K}K_k$ for the kinetic energy operator, and 
\begin{equation}
\label{eq:lin_trott2}
U^\beta_L(\tau)=e^{-i\tau K}e^{-i\tau V}\;,
\end{equation}
for the $\beta$ splitting. Higher order expressions with better error bounds can also be obtained (see Eq.~\eqref{eq:symm_prod_form} and discussion in Appendix~\ref{app:tstep_details}). Here we recall only the 2${}^{\rm nd}$ order expansions that are mostly employ in this work: for the $\alpha$ splitting these are 
\begin{equation}
\label{eq:2nd_order_alpha}
S_\alpha(\tau) = e^{-i\frac{\tau}{2} V}\prod_{k=1}^{N_K}e^{-i\frac{\tau}{2} K_k}\prod_{k=N_K}^{1}e^{-i\frac{\tau}{2} K_k}e^{-i\frac{\tau}{2} V}
\end{equation}
while for the $\beta$ splitting we consider the two options
\begin{equation}
\label{eq:2nd_order_beta1}
S^{K+V}_\beta(\tau) = e^{-i\frac{\tau}{2} K}e^{-i\tau V}e^{-i\frac{\tau}{2} K}
\end{equation}
and
\begin{equation}
\label{eq:2nd_order_beta2}
S^{V+K}_\beta(\tau) = e^{-i\frac{\tau}{2} V}e^{-i\tau K}e^{-i\frac{\tau}{2} V}\;,
\end{equation}
whose implementation requires almost the same number of quantum gates whenever the number of intervals is large (cf. discussion in~\cite{kivlichan2019improved}).

In general the error on these type of product formulae depends on the commutator between the different terms in the sum defining the Hamiltonian~\cite{Trotter59}. One can, however, obtain a rigorous (but not very tight) upperbound on the total error in Eq.~\eqref{eq:approx_time_ev} using only the norms of those operators (see eg.~\cite{Berry2007,Childs2018}).

Using the analytical bounds given in Eq.~\eqref{eq:an_bnd_lin} and Eq.~\eqref{eq:an_bnd_quad} for the linear and symmetric higher order formulae we have estimated the number of segment (Trotter steps) needed to achieve an energy error $\epsilon_\tau = \delta_\tau/\tau$ equal to half the frequency resolution $\Delta \omega$ for two different values of the total time interval $\tau$: the base time $\tau_{base}=2\pi/\Delta H$ (black and green lines) and the whole sequence of $W$ evolutions for a total time of $\tau_{tot}=(2^W-1)*\tau_{base}$ (red and blue lines) where the number of ancilla qubits $W$ is obtained for a fixed resolution $\Delta\omega$ as
\begin{equation}
\label{eq:num_anc}
W = \left\lceil \frac{\Delta H}{\Delta\omega}\right\rceil\;.
\end{equation}

\begin{figure}
    \centering
    \includegraphics[scale=0.33]{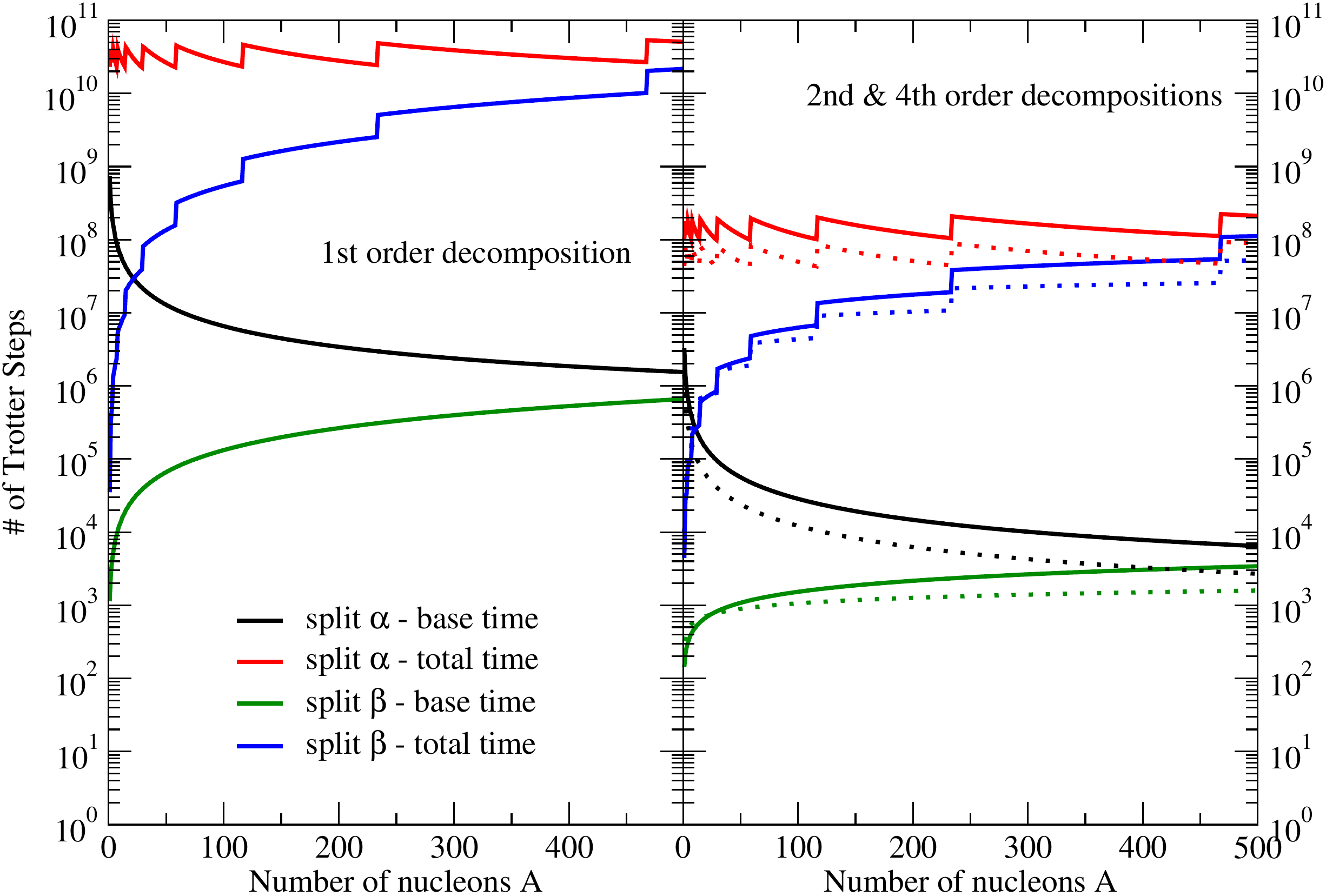}
    \caption{Estimated number of Trotter steps for both splittings of the Hamiltonian and target resolutions $\delta\omega=10$ MeV for Trotter-Suzuki formulas of different order. The left panel shows the linear formulas Eq.~\eqref{eq:lin_trott} and Eq.~\eqref{eq:lin_trott2}, the right panel shows results for both a second order formula (solid lines) and a fourth order one (dotted lines). }
    \label{fig:quad_tsteps}
\end{figure}

We present in Fig.~\ref{fig:quad_tsteps} the results obtained for both splitting methods at the target accuracy $\Delta\omega=10$ MeV (for lower accuracy the difference between 2nd and 4th order formulas is much reduced).

We see that the fourth order formulas (represented as dotted lines in Fig.~\ref{fig:quad_tsteps}) provide an advantage only for big enough problem instances: for the better performing $\beta$ splitting for instance the 4${}^{\rm th}$ order formula becomes advantageous after $A=24$ whereas for lower target accuracy $\delta\omega=100$~MeV (not shown) the break-even point is shifted to $A=234$.

Apart from their dependence on the norms of the Hamiltonian terms instead of their commutators, an important deficiency of the bounds used above is also their inability to differentiate between different ordering of operators in higher order formulae. We provide a more detailed discussion on the derivation of commutator bounds in the Appendix~\ref{app:tstep_details}. Here, in Fig.~\ref{fig:second_comm_tsteps}, we show only the effect on the more efficient 2${}^{\rm nd}$ product formulae Eq.~\eqref{eq:2nd_order_alpha}, Eq.~\eqref{eq:2nd_order_beta1}, and Eq.~\eqref{eq:2nd_order_beta2}. For all curves the target accuracy was fixed to $\Delta\omega=100$~MeV for the two splitting methods.

\begin{figure}
    \centering
    \includegraphics[scale=0.33]{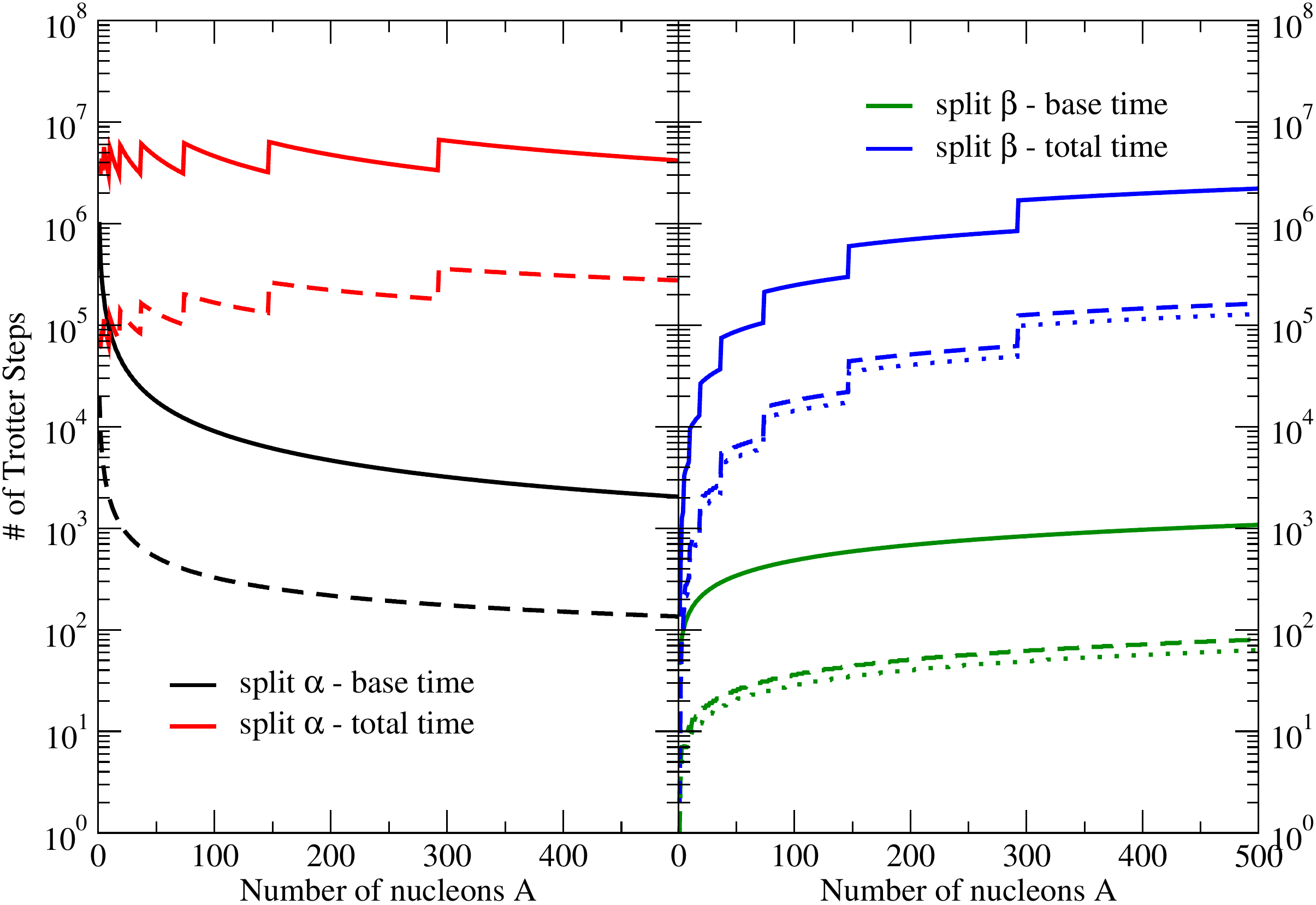}
    \caption{Comparison of the analytical (solid lines) vs. commutator bounds (dashed lines) $r_{2:A}$ and $r_{2:C}$  for the quadratic Trotter-Suzuki breakup with both splitting schemes ($\alpha$ for the left panel and $\beta$ for the right one) at fixed target resolution $\Delta\omega=100$ MeV. The dashed lines for the $\beta$-splitting corresponds to the choice $(V+K)$ while the dotted lines for the complementary choice $(K+V)$ (see text for details).}
    \label{fig:second_comm_tsteps}
\end{figure}

We turn now into a more detailed discussion on the computational cost for a single time step needed to implement the time propagator using different implementation strategies. For both splitting methods we will need to design three different unitary operators (more details in Appendix~\ref{app:tstep_impl})
\begin{equation*}
U_1(\tau)=e^{-i\tau V}\;\; U_2(\tau)=e^{-i\tau K}\;\; U_3(\tau)=\prod_{k=1}^{N_K}e^{-i\tau K_k}\;.
\end{equation*}
In our derivation we will consider the connectivity of qubits to follow a 2D square lattice topology and, even under this constraint, the implementation of the diagonal unitary $U_1$ is relatively simple (see Appendix~\ref{app:u1imp}).

Due to it's (mild) non-locality, the most expensive term to implement is the hopping term. Depending on the splitting scheme, we will adopt (similarly to the approach described in \cite{kivlichan2019improved}) the fast fermionic Fourier transform (FFFT) algorithm~\cite{Verstraete2009} (or it's variants~\cite{Wecker2015,Kivlichan2018}) for the implementation of splitting $\beta$ while employ a fermionic-SWAP network~\cite{Kivlichan2018} to implement the product of unitaries needed for the splitting $\alpha$.

\begin{figure}
    \centering
    \includegraphics[scale=0.33]{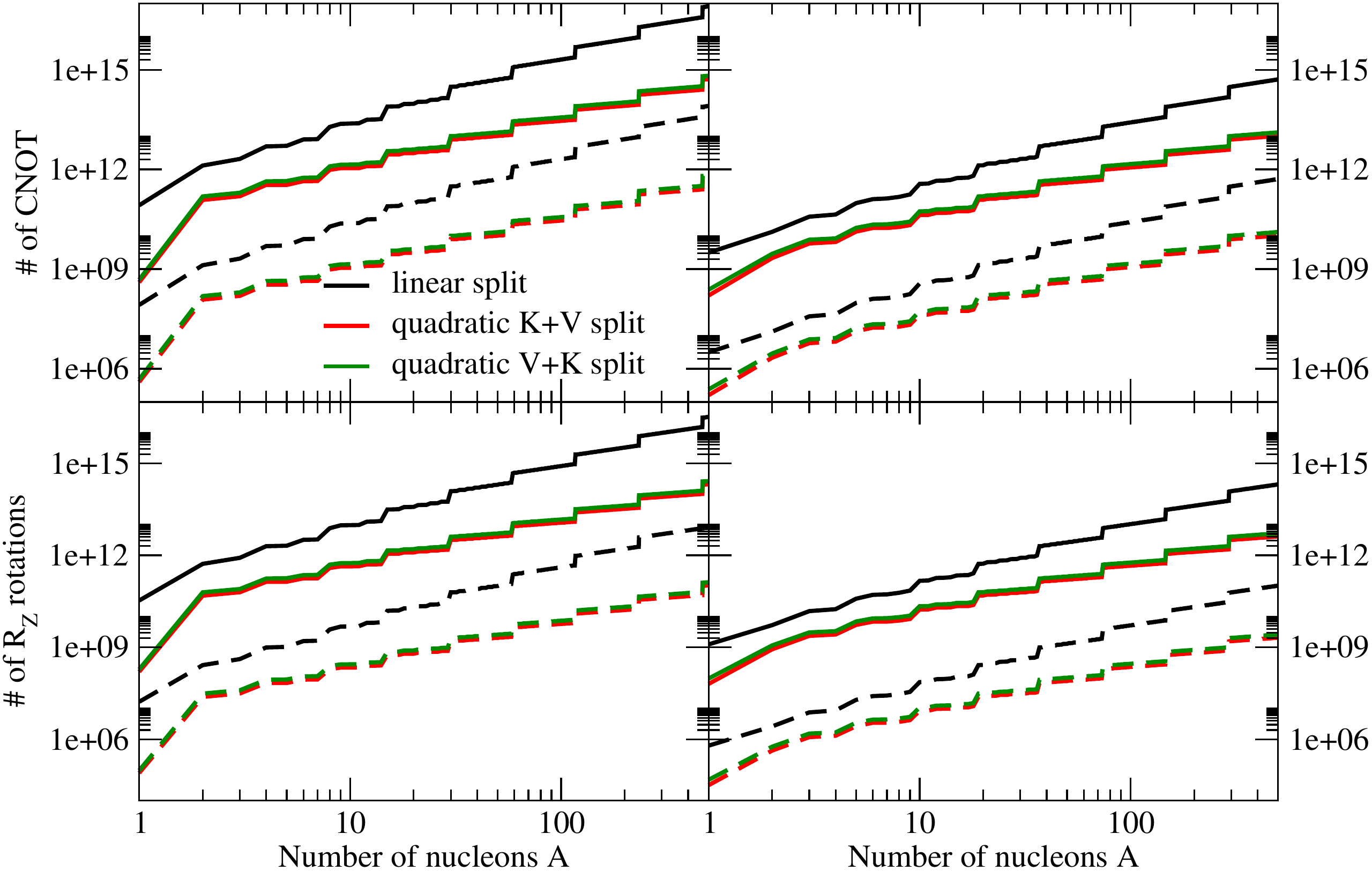}
    \caption{Estimated gate count, in the CNOT$+R_z$ basis, of the phase estimation kernel of the linear response algorithm of \cite{Roggero2018} as a function of nucleon number. Shown are results for the $\beta$ splitting and both first and second order Trotter decompositions. The left panel is for an energy resolution $\Delta\omega=10$ MeV and the right for $\Delta\omega=100$ MeV. The solid lines correspond to the serial execution while parallelism is exploited for the dashed ones.}
    \label{fig:gcount_beta}
\end{figure}

Results of the cost estimates for a realistic system with $M=10^3$ and $N_f=4$ are presented in Fig.~\ref{fig:gcount_beta}. In this setup, performing the calculation for $^{40}$Ar would require $~4012$ qubits ($\pm4$ depending on target resolution and the particular implementation) and $\sim10^{10}$ CNOT and $\sim10^9$ rotation for the higher resolution $\Delta\omega=10$ MeV and $\sim5\times10^8$ CNOT and $\sim10^8$ rotations at lower resolution $\Delta\omega=100$ MeV. These estimates put a full computation of neutrino scattering off Ar at the same complexity level as factorizing a 1024-bit integer (cf.~\cite{Kutin2006,Childs2018}) and possibly out of reach to near term NISQ devices. In the next section we explore possible improvements to this estimate using qubitization.

\subsection{Different scheme using qubitization}
\label{sec:new_scheme_qubit}
Here we present an alternative strategy to perform a variant of the LR algorithm from \cite{Roggero2018} which doesn't require the use of the time-evolution unitary but only of a quantum walk operator known as qubiterate~\cite{Low_and_Chuang_2019}. One possible definition of this unitary, acting both on the system register and an additional register of ancilla qubits, is
\begin{equation}
\label{eq:qubiterate}
Q = e^{i \arcsin\left(H/\lambda\right)R_Y}
\end{equation}
where $\lambda\geq\|H\|_1$ is a scaling factor needed to ensure the argument of the arcsin has norm bounded by one and, for every eigenvector of $H$, the operation $R_Y$ acts non-trivially only on a 2-dimensional subspace of the ancilla register's Hilbert space (see Appendix~\ref{app:qubitiz} for a more detailed exposition). This unitary operator can be implemented exactly using qubitization (see \cite{Low_and_Chuang_2019} and the discussion in Appendix~\ref{app:qubitiz}) which exploits the following decomposition (apart from a global phase) in terms of two basic unitaries $V_P$ and $V_S$ and a reflection
\begin{equation}
\label{eq:qubit_decomp}
Q = V_P^\dagger \Pi_0 V_P V_S
\end{equation}
where $\Pi_0 = \left(\rvert0\rangle\langle0\lvert-\mathbb{1}\right)$ is a reflection around $\ket{0}$, the operation $V_P$ is called the {\it prepare} and $V_S$ the {\it select} unitary (see Eqs.~\eqref{eq:select_def} and \eqref{eq:prepare} in Appendix~\ref{app:qubitiz}).

Since the spectra of $U(t)$ and the qubiterate of Eq.~\eqref{eq:qubiterate} are similar, the idea (originally proposed in~\cite{Poulin2018} and~\cite{Babbush2018}) is now to use the exact qubiterate for doing phase estimation instead of the (approximate) time evolution operator. The first main difference is that, due to the rescaling, the number of ancilla qubits used for phase estimation (or equivalently the number of applications of the qubiterate) will need to increase accordingly. In particular we have, for target precision $\Delta\omega$, the result
\begin{equation}
W_q = \left\lceil \log_2\left(\frac{\lambda}{\Delta\omega}\right)\right\rceil \sim \log_2\left(\frac{\lambda}{\Delta H}\right) W
\end{equation}
where $W$ was the previous result for the qubit count using time evolution. The second main difference is that in order to obtain the final state of the scattering process we need to perform a rotation from the eigenvectors of the qubiterate to those of $H$, one way to do this is to use the strategy proposed in~\cite{Poulin2018}.
\begin{figure}
    \centering
    \includegraphics[scale=0.33]{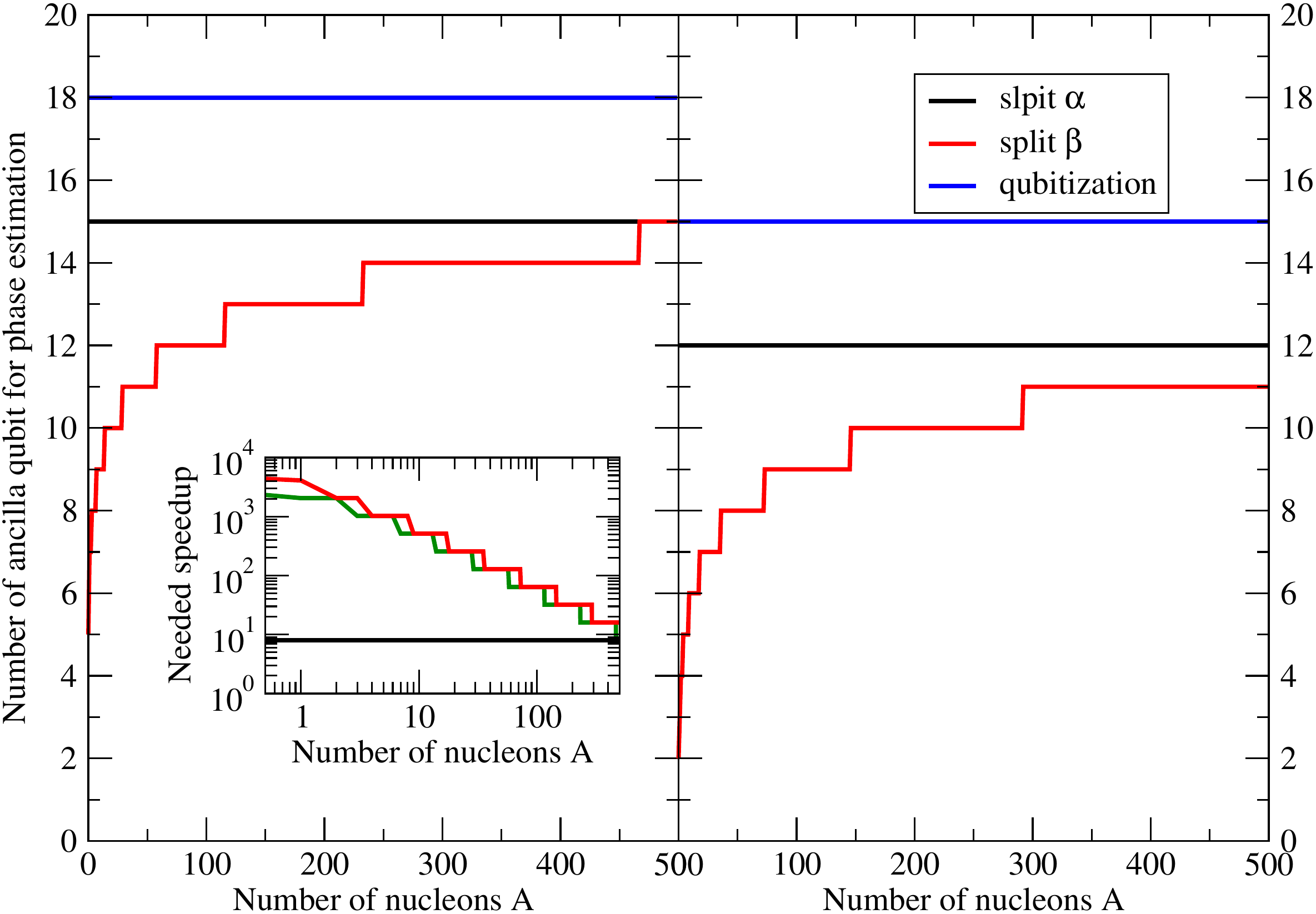}
    \caption{Estimated number of ancilla qubit needed for a fixed target precision of $\Delta\omega=10$ MeV (left panel) and $100$ MeV (right panel) using either time evolution with both split methods described in the main text and qubitization. The inset shows the needed speed up the qubiterate needs to show with respect to the time evolution circuit for the base time $\tau=2\pi/\Delta H$ in order to be competitive.}
    \label{fig:anc_with_qub}
\end{figure}

In Fig.~\ref{fig:anc_with_qub} we show the expected increase as a function of the number of nucleons for two different target accuracies: $\Delta\omega=10$ MeV in the left panel and $\Delta\omega=100$ MeV for the right one. In the inset we show the ratio between the number of applications of the qubiterate vs the number of applications of the base time evolution $U(\tau)$ for time $\tau=2\pi/\Delta H$, it represents the needed speedup in gate count of the qubiterate with respect to $U(\tau)$ for the qubitization strategy to be worth it. As expected the difference between the $\alpha$-split method and qubitization do not depend on the particle number and the ratio is stable at $8$ (meaning the implementation of Eq.~\eqref{eq:qubiterate} can require up to $8$ times more gates than time evolution as shown by the black line in the inset). For the $\beta$-split scheme, this ratio is 128 in the $^{40}$Ar region. 

In order to employ the qubiterate $Q$ for the QPE part of the algorithm we need to implement the operation ${}_cQ$ controlled on an ancilla. Using the decomposition presented above in Eq.~\eqref{eq:qubit_decomp} we can write
\begin{equation}
{}_cQ = {}_cV_P^\dagger {}_c\Pi_0{}_cV_P{}_cV_S=V_P^\dagger {}_c\Pi_0V_P{}_cV_S\;,
\end{equation}
where in the second equality we removed the controls on the prepare (this simplification was proposed before, see for instance~\cite{Childs2018}). One can simplify this further and drop the control on the {\it select} unitary if we choose to define $V_S\rvert0\rangle=\rvert0\rangle$ when acting on the $\ket{0}$ state of the ancilla register and perform an initial controlled-{\it prepare} when initializing the ancilla system.

\begin{figure}
    \centering
    \includegraphics[scale=0.3]{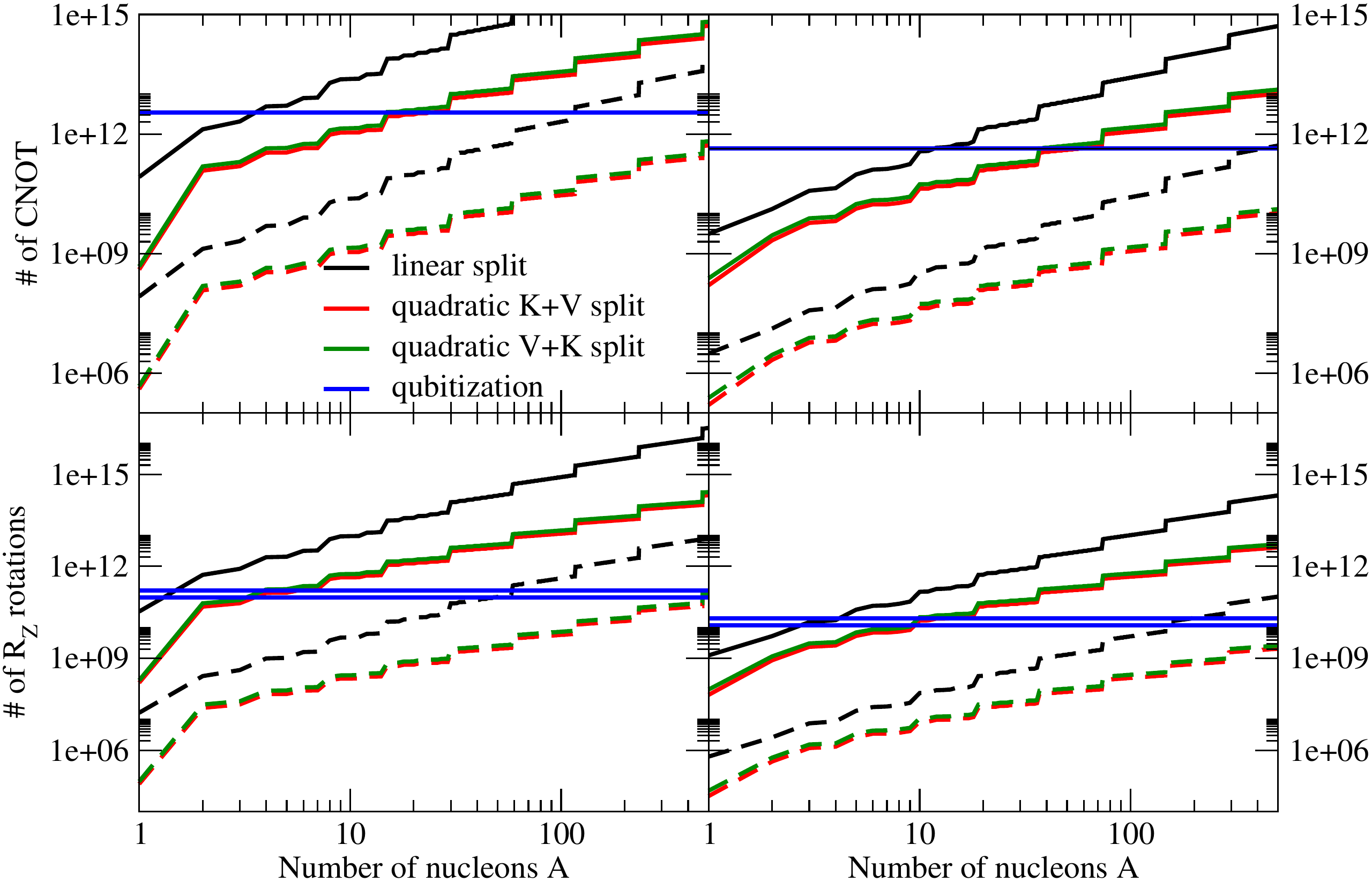}
    \caption{Estimated total gate count in CNOT + $R_Z$ basis for the $\beta$ splitting and the variant based on the qubiterate. \label{fig:total_cost_woth_qub}}
\end{figure}

Using the implementation proposed in \cite{Childs2018} (and presented in more detail in Appendix~\ref{app:qub_cost} for completeness) we found the resource estimates reported in Fig.~\ref{fig:total_cost_woth_qub}. In these results we considered only the cost for implementing the {\it prepare} unitary $V_P$ together with the control circuit of the {\it select} unitary $V_S$ and are therefore lower bounds on the resource cost. We then see that, even though this methodology has optimal asymptotic scaling~\cite{Low_and_Chuang_2019}, the inherent costs of implementing qubitization is already expensive enough to lose the competition with the parallel circuits devised above. 

A possible way to reduce the gate count needed to successfully perform a calculation of the response of $^{40}$Ar would be to exploit stochastic algorithms like the recently proposed qDRIFT~\cite{Campbell2019}. We plan to explore this possibility in future work. 

\section{Triton toy model}
\label{sec:actual_implementation}
For a simplified problem on present-day computers,
we consider a system of 3 nucleons on a 2x2 lattice
with periodic boundary conditions.  One of the nucleons
is chosen to be static (infinite mass)
on a specific lattice site. This can be thought of as
a triton (a nucleus with one protons and two neutrons),
or the static nucleon can be thought of as providing
a static field in which the interacting pair propagates.

Calculations of realistic response demonstrate that two-nucleon physics incorporates much of the information about nuclear response\cite{Pastore2019}, making even such a simple problem important.  
The fixed particle is ultimately a 
source of additional final state scattering which in traditional event generators is included as a semi-classical evolution.  Quantum computers will eventually be able to treat the full problem for A nucleons quantum mechanically. In the near term these kinds of
models allow for tests
of the generator paradigm, where at the vertex a struck nucleon 
or nucleon pair is treated quantum mechanically and then propagates 
through the rest of the nucleus in a semiclassical manner.

The Hamiltonian we use is:
\begin{equation}
\begin{split}
H&= -t \sum_{f=1}^{N_f}\sum_{\langle i,j\rangle}  c^\dagger_{i,f}c_{j,f} + 2dtA \\
&+U \sum_{i=1} \sum_{f< f'}^{N_f} n_{i,f}n_{i,f'} + V \sum_{f< f'< f''}^{N_f}\sum_{i=1} n_{i,f}n_{i,f'}n_{i,f''}\\
&+U \sum_{f=1}^{N_f} n_{1,f} + V \sum_{f< f'}^{N_f} n_{1,f}n_{1,f'}
\end{split}
\label{eq:lattice_joe}
\end{equation}
where the static nucleon is placed on lattice site 1.

For this example we use only 2 dynamical particles and we set $N_f=2$. 
On a $2 \times 2$ lattice with $N_f=2$ modes we find that the
$2 \times 2$ Hamiltonian in second quantization
with the simple Jordan-Wigner mapping described above (1 qubit for each single-particle orbital) will require a total of 8 qubits to encode the problem. We are, however, interested in the sector containing $A=2$ dynamical particles whose dimension is only $16$ and should require just $4$ qubits. In the following we will use a first-quantized mapping that accomplishes this minimal encoding.

\begin{figure}
    \centering
    \includegraphics[scale=0.33]{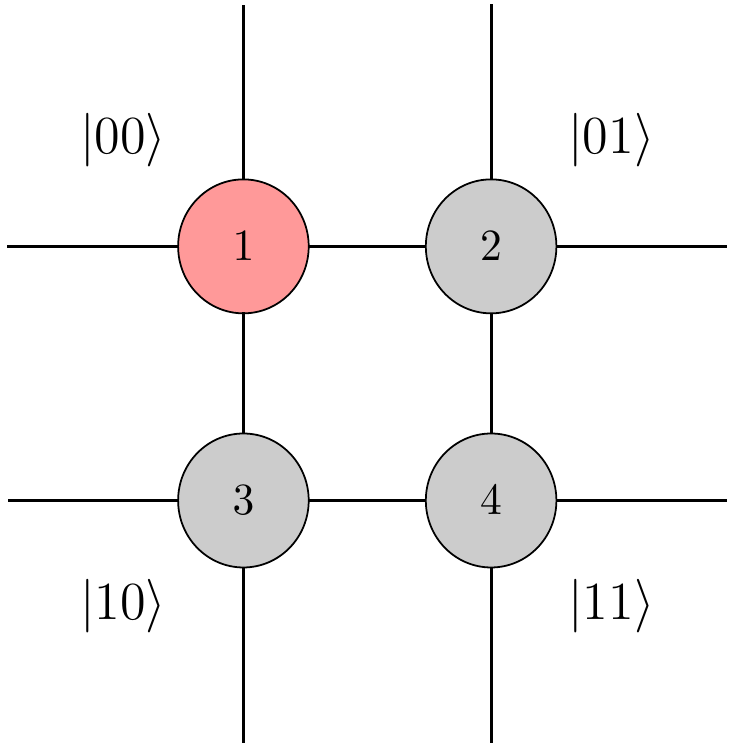}
    \caption{Qubit mapping for a single fermion. \label{fig:qmap}}
\end{figure}

We can use 2 qubits per particle to store its lattice location in the following way (see also Fig.~\ref{fig:qmap})
\begin{equation}
\ket{1}\equiv\ket{\downarrow\downarrow}\quad\ket{2}\equiv\ket{\downarrow\uparrow}\quad\ket{3}\equiv\ket{\uparrow\downarrow}\quad\ket{4}\equiv\ket{\uparrow\uparrow}\;.
\end{equation}

%\begin{tikzpicture}
%\begin{center}
%\draw[step=2cm,gray,very thin] (0.1,0.1) grid (5.9,5.9);
%\node[circle,draw=black,fill=white!60!red,minimum size=30] at (2,4) {1};
%\node[above left] at (1.5,4.5) {\large$\rvert 0 0 \rangle$};
%\node[circle,draw=black,fill=white!80!black,minimum size=30] at (4,4) {2};
%\node[above right] at (4.5,4.5) {\large$\rvert 0 1 \rangle$};
%\node[circle,draw=black,fill=white!80!black,minimum size=30] at (2,2) {3};
%\node[below left] at (1.5,1.5) {\large$\rvert 1 0 \rangle$};%$\bigg\rvert$\large{$1 0$}$\bigg\rangle$};
%\node[circle,draw=black,fill=white!80!black,minimum size=30] at (4,2) {4};
%\node[below right] at (4.5,1.5) {\large$\rvert 1 1 \rangle$};
%\end{center}
%\end{tikzpicture}

The hopping term in the kinetic energy is very simple and takes the form
\begin{equation}
H_{hop} = H^A_{hop}\otimes\mathbf{1}_{B} + \mathbf{1}_{A}\otimes H^B_{hop}
\end{equation}
where
\begin{equation}
H^A_{hop} = -2t \begin{pmatrix} 0 & 1 &1 &0 \\
1 & 0 &0 &1 \\
1 & 0 &0 &1 \\
0 & 1 &1 &0 \\
\end{pmatrix} \equiv -2t \left(X_1\otimes\mathbf{1}_{2} + \mathbf{1}_1\otimes X_{2} \right)
\end{equation}
where $X_k$ is the Pauli-X operator applied to qubit $k$ and the additional factor of 2 comes from the periodic boundary conditions.
The total hopping term reads then
\begin{equation}
H_{hop} = -2t \left(X_1 +  X_{2} + X_3 + X_4 \right)
\end{equation}
where we dropped the identity operators for simplicity.

For the diagonal part, we can extract an overall piece proportional to the identity on all qubits with coefficient $8t+U$; to change the diagonal element corresponding to the state $\ket{11}$ we add $2U+V$; and when both particles are on different lattice sites and neither of which is 1 by adding $-U$.

The procedure to do this in terms of Pauli operators is very simple as shown by a couple of examples. Consider the two sets of operators
\begin{equation}
M_k = \frac{\mathbb{1}_k-Z_k}{2}\quad \Pi_k = \frac{\mathbb{1}_k+Z_k}{2}\;.
\end{equation}
In terms of these operators we have
\begin{equation}
\begin{split}
(2U+V)&\ket{11}\bra{11}=(2U+V)\ket{\downarrow\downarrow\downarrow\downarrow}\bra{\downarrow\downarrow\downarrow\downarrow}\\
=&(2U+V)\left[\Pi_1\otimes \Pi_2\otimes \Pi_3\otimes \Pi_4\right]
\end{split}`
\end{equation}
and
\begin{equation}
\begin{split}
-U&\ket{23}\bra{23}=-U\ket{\downarrow\uparrow\uparrow\downarrow}\bra{\downarrow\uparrow\uparrow\downarrow}\\
 =&-U\left[\Pi_1\otimes M_2\otimes M_3\otimes \Pi_4\right]
\end{split}
\end{equation}
and so on for the other terms. 

The limiting case, $V=-4U$, results in the following simplified Hamiltonian:
\begin{equation}
\label{eq:red_ham}
\begin{split}
H&=8t+\frac{U}{2} - 2t \sum_{k=1}^4 X_k \\
&-\frac{U}{4} \left(Z_1Z_4 + Z_2Z_3\right)-\frac{U}{4} \sum_{i<j<k} Z_iZ_jZ_k\;.
\end{split}
\end{equation}
This choice of parameters is motivated by the requirement that the 3-body repulsive term be larger than the 3 pair interaction energy in order to prevent the collapse of the bound state.
In the following we will consider the following numerical values: $t=1.0$, $U=-7.0$, $V=28$.

\subsection{State preparation}
A simple trial state that is also economic to optimize can be obtained by considering the following circuit
\begin{equation}
\Qcircuit @C=0.8em @R=.7em {
&\gate{R_y(\theta)}&\ctrl{3}&\qw     &\qw               &\ctrl{3}&\qw&\qw\\
&\gate{R_y(\theta)}&\qw     &\ctrl{1}&\qw               &\qw     &\ctrl{1}&\qw \\
&\gate{R_y(\theta)}&\qw     &\ctrl{0}&\gate{R_y(\phi)}&\qw     &\ctrl{0}&\qw\\
&\gate{R_y(\theta)}&\ctrl{0}&\qw     &\gate{R_y(\phi)}&\ctrl{0}&\qw     &\qw
}\;,
\end{equation}
parametrized by two angles $(\theta,\phi)$ and requires only linear connectivity to be implemented.

The entanglement structure is inspired by the CCSD-type wavefunction that we would construct in the absence of the $3-$body terms in the Hamiltonian of Eq.~\eqref{eq:red_ham} and by the fact that the Hamiltonian is real in the computational basis.

As can be seen from Tab.~\ref{tab:trial_state}, despite its simplicity this trial state has only about $10\%$ error in the energy and sum rules are comparable with the exact ground state. In the results presented in this work, the optimization of the two parameters of our trial state is performed off-line using a simulator locally. After extensive experimentation we, in fact, determined that this was the most efficient and accurate strategy: this is possibly a consequence of the simplicity of the problem. In  the central two rows of Tab.~\ref{tab:trial_state}, we present the results obtained by estimating the properties of the state generated on a real quantum processor. In particular, we mapped our four computational qubits into qubit $5$,$0$,$1$ and $6$ respectively on the IBMQ 20 qubit machine Poughkeepsie~\cite{IBMQ_Pough}. In the first line denoted 'QPU bare', we report the bare result obtained from a statistical analysis of $324$ runs each comprising of $8192$ repetitions (shots) but without performing any form of error mitigation. The next line shows the much better result obtained by mitigating both read-out noise and the decoherence effect coming from the CNOT gates (see Section~\ref{subs:error_mitigation} for more details). 

\begin{table}[]
\begin{tabular}{l|cccc}
            & Energy & S(0,1) & S(1,0) & S(1,1) \\ \hline
exact g.s.  & -4.843 & 2.038 & 2.038 & 2.054 \\
trial state & -4.415 & 2.024 & 2.024 & 2.366\\ \hline
QPU bare  & -2.645(15) & 2.0290(23) & 2.0242(24) & 2.1572(25) \\
QPU corr  & -4.4187(98) & 1.9993(35) & 1.9926(36) & 2.2789(51) \\ \hline
QPU sym  & -4.322(33) & 2.0105(69) & 2.0030(45) & 2.3341(95) \\
%QPU sym [0-err]  & -4.268(55) & 2.0082(42) & 2.0078(69) & 2.324(15) \\
\end{tabular}
\caption{Results for the ground state energy and the static structure factor. Errors in the experimental result account for statistical fluctuations only. \label{tab:trial_state}}
\end{table}

In the last line of Tab.~\ref{tab:trial_state} we report instead the (error mitigated) results obtained from $108$ runs using a more symmetric version of the trial state above and shown below
\begin{equation}
\Qcircuit @C=0.8em @R=.7em {
&\gate{R_y(\theta)}&\ctrl{3}&\qw     &\gate{R_y(\phi/2)}&\ctrl{3}&\qw&\qw\\
&\gate{R_y(\theta)}&\qw     &\ctrl{1}&\gate{R_y(\phi/2)}&\qw     &\ctrl{1}&\qw \\
&\gate{R_y(\theta)}&\qw     &\ctrl{0}&\gate{R_y(\phi/2)}&\qw     &\ctrl{0}&\qw\\
&\gate{R_y(\theta)}&\ctrl{0}&\qw     &\gate{R_y(\phi/2)}&\ctrl{0}&\qw     &\qw
}\;.
\end{equation}

The added symmetry seems to bring some advantage in the $(1,1)$ sum rule but the added noise caused by additional noisy rotations seems to be detrimental for the energy.

\subsection{Real time dynamics}
In the general case ($V\neq0$ and $V\neq -4U$) one can use the result from \cite{Welch2014} which implies that we would need $14$ CNOT and $15$ single qubit rotations for the diagonal part of the propagator plus $4$ more $X$ rotations for the hopping term resulting in $14$ CNOT and $19$ rotations (with more constraints like having a circle topology this can increase to $16$ CNOT. See also Eq.~\eqref{eq:optimal_pot_circuit}).
For the special case $V=-4U$ a simpler expression can be found
\begin{widetext}
\begin{equation}
\Qcircuit @C=1em @R=.7em {
\lstick{q0}&\ctrl{3}&\qw     &\qw                 &\ctrl{2}&\qw     &\qw                 &\qw     &\qw                 &\ctrl{3}&\qw     &\qw                  &\qw     &\qw     &\gate{R_x(\theta_1)}&\qw\\
\lstick{q1}&\qw     &\ctrl{1}&\qw                 &\qw     &\ctrl{2}&\qw                 &\qw     &\qw                 &\qw     &\ctrl{2}&\qw                  &\qw     &\ctrl{1}&\gate{R_x(\theta_1)}&\qw\\
\lstick{q2}&\targ   &\targ   &\gate{R_z(\theta_2)}&\targ   &\qw     &\gate{R_z(\theta_2)}&\ctrl{1}&\qw                 &\qw     &\qw     &\qw                  &\ctrl{1}&\targ   &\gate{R_x(\theta_1)}&\qw\\
\lstick{q3}&\targ   &\qw     &\gate{R_z(\theta_2)}&\qw     &\targ   &\gate{R_z(\theta_2)}&\targ   &\gate{R_z(\theta_2)}&\targ   &\targ   & \gate{R_z(\theta_2)}&\targ   &\qw     &\gate{R_x(\theta_1)}&\qw\\
}
\end{equation}
\end{widetext}
with $\theta_1=4t\tau$ and $\theta_2=\tau U/2$. This implementation requires $10$ rotations and $10$ CNOT. The problem with this expression is that it requires entangling gates on all but one pair of qubits (ie. in the expression above there is no connection $q_0 \leftrightarrow q_1$ but all others).

With the additional connectivity constraints of the IBM QPU 'Poughkeepsie' we found the following circuit
\begin{widetext}
\begin{equation}
\Qcircuit  @C=.7em @R=.3em @!R {
\lstick{q0}&\qw      &\targ    &\gate{R_z(\theta_2)}&\targ    &\qw      &\targ    &\ctrl{1}&\gate{R_z(\theta_2)}&\ctrl{1}&\ctrl{3}&\qw   &\targ   &\gate{R_z(\theta_2)}&\targ&\qw&\targ&\ctrl{1}&\qw                  &\ctrl{1}&\qw        &\gate{R_x(\theta_1)}&\qw\\
\lstick{q1}&\targ    &\ctrl{-1}&\gate{R_z(\theta_2)}&\ctrl{-1}&\targ    &\qw      &\targ   &\gate{R_z(\theta_2)}&\targ   &\qw     &\qswap &\ctrl{-1}&\qw        &\ctrl{-1}&\targ&\qw&\targ&\gate{R_z(\theta_2)}&\targ&\targ         &\gate{R_x(\theta_1)}&\qw\\
\lstick{q2}&\ctrl{-1}&\qw      &\qw                 &\qw      &\ctrl{-1}&\qw      &\qw     &\qw                 &\qw     &\qw     &\qswap\qwx &\qw     &\qw         &\qw&\ctrl{-1}&\qw &\qw&\qw                      &\qw&\ctrl{-1}                               &\gate{R_x(\theta_1)}&\qw\\
\lstick{q3}&\qw      &\qw      &\qw                 &\qw      &\qw      &\ctrl{-3}&\qw     &\qw                 &\qw     &\targ   &\qw   &\qw     &\qw           &\qw&\qw&\ctrl{-3}&\qw&\qw                    &\qw&\qw                                     &\gate{R_x(\theta_1)}&\qw \gategroup{1}{11}{4}{12}{.7em}{--}
}\;,
\end{equation}
\end{widetext}
where in the box denoted with the dashed line we perform a swap of both $q1\leftrightarrow q2$ and $q0\leftrightarrow q3$. Of the latter two of the three CNOT involved in the operation cancel with neighboring gates.

\begin{figure}
    \centering
    \includegraphics[scale=0.33]{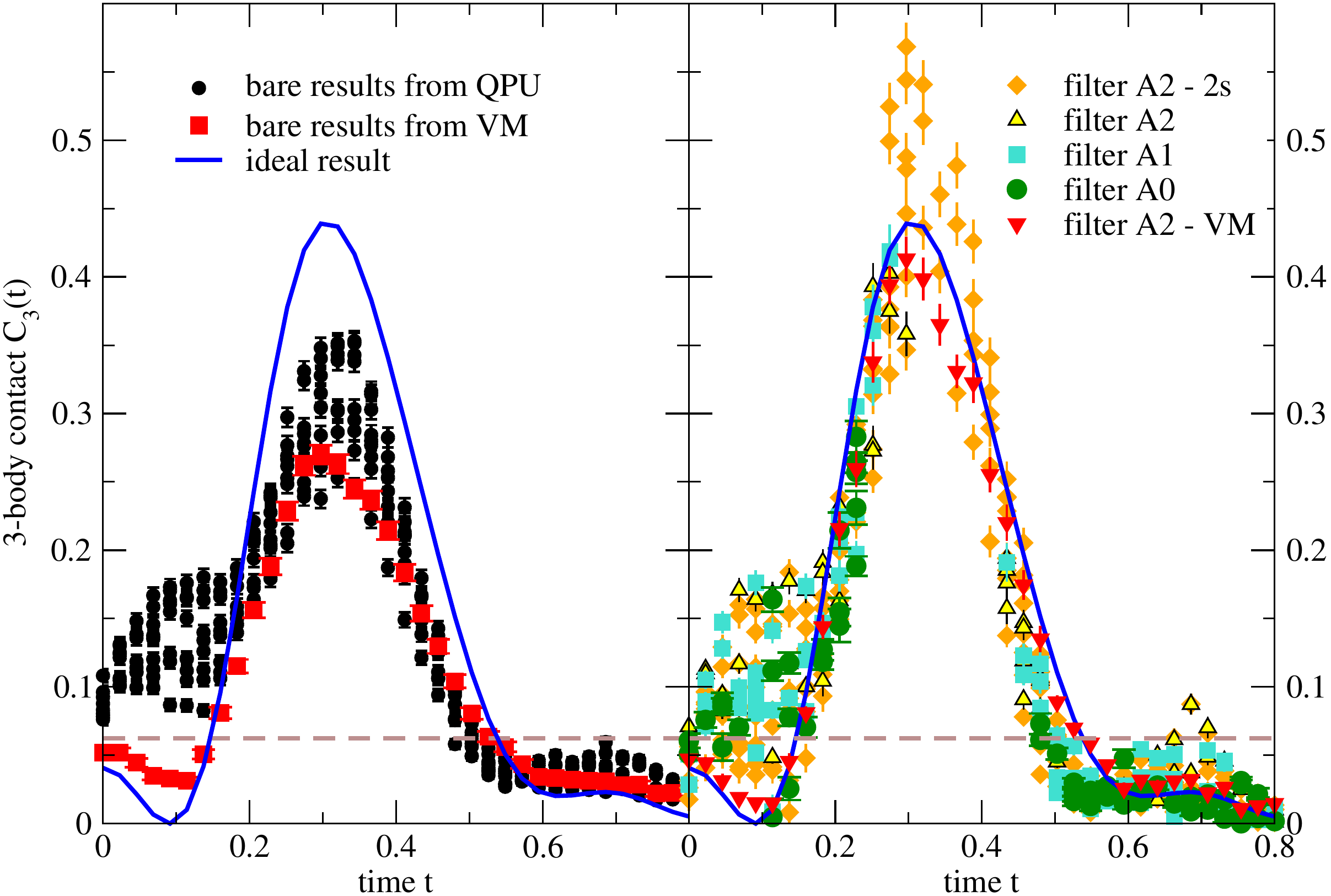}
    \caption{Probability of finding the 3 nucleons on the same site as a function of time using both a real QPU (black circles) and a simulated VM (red squares). See text for a description of the left panel.\label{fig:d0oft_cmp}}
\end{figure}

We can now show results for some dynamical property. In Fig.~\ref{fig:d0oft_cmp} we plot the 3-body contact density
\begin{equation}
\label{eq:3bcontact}
C_3(t) = \langle \Psi(t) \lvert \Pi_{0000} \rvert \Psi(t)\rangle \equiv \lvert \langle 0000\vert\Psi(t)\rangle\rvert^2
\end{equation}
as a function of time starting at time $t=0$ with the trial state of the previous section. The expression above measures the probability of the three nucleons to be on the same site (the state $\ket{0000}$ in our basis). The time evolution is obtained by means of the linear Trotter decomposition described above and therefore starts to deviate considerably from the exact time evolution at around $t\sim0.04$.

In the left panel we show, together with  the exact result with the solid blue line, the bare results obtained by running the algorithm on either the actual quantum device (black circles) or on a local virtual machine employing a noise model designed to mimic the behaviour of the real device (red squares) \cite{Qiskit}. The hardware results were obtained using the 'Poughkeepsie' QPU (backend version 1.2.0) over a 3 weeks period starting on 23 August 2019 and adopting the mapping $(q_0,q_1,q_2,q_3)\to(q_5,q_0,q_1,q_6)$ from the 4 logical qubits to the hardware ones. The corresponding results on the Virtual Machine used the noise model configured with the calibration data on 11 September 2019.

In both cases, we see that the results tend to relax towards the classical completely depolarized value of $1/16$ (dashed brown line) but that we can still detect a reasonable signal. The observed large bias at small times might be attributable to control errors in the device and unfortunately does not allow this particular set of qubits to be used to perform multiple Trotter steps as the error in the useful region is too large. Different choices for the logical to physical qubit mapping can improve the fidelity in the small time region. 

We want turn our attention to the right plot in Fig.~\ref{fig:d0oft_cmp}. As explained in more detail in the next subsection, we have attempted to mitigate the systematic errors caused by hardware noise by performing 3 independent noise extrapolations and comparing them to assess both the stability of our extrapolations and the stability of the machine during a particular run. Whenever the different schemes do not agree we increase an error counter and filter the final results using the total error count as a metric for the run quality. In the right panel of Fig.~\ref{fig:d0oft_cmp} we present the result after this mitigation procedure for different values of the error count starting from 0 (filter $A0$ in the figure) up to 2 errors (filter $A2$). In addition to the results obtained on hardware with this approach, we also plot the results at the 2 error level of accuracy for both the synthetic data produced by the VM (the red squares on the left panel) and the results obtained by relaxing the consistency checks at the $2\sigma$ level of precision. We can see that the simpler noise model implemented in the VM can be completely mitigated using this strategy while for the real hardware case there seems to be a problem in the time region $t\in[0.3-0.4]$ where no results with good enough quality can be obtained. We will provide a possible explanation for this phenomenon after discussing in more details the mitigation procedure adopted in our work.

\subsection{Error Mitigation}
\label{subs:error_mitigation}
In our final results like those shown in Fig.~\ref{fig:d0oft_cmp} we perform two types of error mitigation: a read-out correction on the measured distributions and a noise extrapolation assuming that the dominant noise channel is the one associated with the execution of a CNOT gate (cf.~\cite{Li2017,Dumitrescu2018,Endo2018}). In the future we would also like to investigate the use of twirling (see eg.~\cite{Wallman2016}) to contrast the control errors affecting the small time results shown above.

The correction for measurement errors is obtained by first attempting a simple procedure where we assume errors are qubit-independent and described by a distortion of the two measurement operators
\begin{equation}
\begin{split}
&\rvert\widetilde{0}\rangle\langle\widetilde{0}\lvert = (1-p_0)\rvert0\rangle\langle0\lvert + p_1 \rvert1\rangle\langle1\lvert\\
&\rvert\widetilde{1}\rangle\langle\widetilde{1}\lvert = p_0 \rvert0\rangle\langle0\lvert + (1-p_1)\rvert1\rangle\langle1\lvert
\end{split}
\end{equation}
and use the results of two calibration measures where we prepare both basis states and perform a Z measurement to obtain the empirical error matrix 
\begin{equation}
N_k = \begin{pmatrix}
1-p_0 & p_1 \\
p_0 & 1-p_1 \\ 
\end{pmatrix} \,.
\end{equation}
Here, the subscript $k$ identifies a particular qubit on the hardware. Noise free results are then obtained by applying the inverse of this matrix to the measured distribution while the errors are propagated correctly in the process (see also \cite{Roggero2019} for more details on the procedure).

If this simple scheme produces an unphysical distribution (with, for example, negative entries), the central value of the corrected distribution is obtained using a least square inversion and a generalized procedure where the calibration is obtained from set of $2^n$ state preparation as implemented in qiskit Ignis~\cite{Qiskit}, while the error is estimated from the simpler procedure used before. In our experience, violations of the independent qubit error model are rather rare. In order to track the quality of an experimental run we will add one error if the simple procedure fails (now the threshold is set to $2\sigma$ for the check).

In order to obtain an estimate of the noise free result, we use the idea of noise amplification and extrapolation used successfully in the past \cite{Endo2018,Dumitrescu2018,Kandala2019}. The idea is simple: imagine a model for the parametric dependence of an observable $M$ to the noise strength $\epsilon$ (for instance a low order polynomial for small $\epsilon$) and we can control the noise strength accurately enough to produce estimates $M(k \times \epsilon)$ at larger error rate (ie. $k>1$), then we could extrapolate the result to the zero error limit.

In this work we employ 3 different strategies together:
\begin{itemize}
    \item Richardson: in the regime where the circuit depth is very small and only a small amount of errors are contaminating the results, it makes sense to look for a Richardson type extrapolation obtained by computing the exact polynomial interpolant of the noisy points we have (cf.~\cite{Temme2017}). At order $3$, as in our case, we will obtain a cubic. The lowest order compatible with the higher ones (at the 1 $\sigma$ level) is considered the preferential one. If a result satisfying this compatibility is found, we increase the error count by one, drop the highest order point and try again. If still no valid point is found, the Richardson extrapolation is deemed failed.
    \item Polynomial: in the same small error regime if the rate is small enough we should be able to fit multiple points with the same low order polynomial (cf.~\cite{Dumitrescu2018}). Here we attempt to perform polynomial fits up to third order of all the (4 in our case) points available. As for the previous method we look for the lowest order fit with $\chi^2\leq1$ and compatible with the higher order fits. Similarly, failure over the $4$ points increases the error count and we try without the highest order point. If the procedure fails the second time, the polynomial fit is deemed failed.
    \item Exponential: when the error rate is sufficiently large (or the gate count is), one could expect the results to decay exponentially towards the fully depolarized state (cf. discussion in \cite{Endo2018,Endo2019}). We attempt a two point exponential fit to the results and, as for the methods above, look for compatibility at higher orders and raise the error count when this cannot be found.
\end{itemize}

A run is considered to have been executed successfully when at least one technique produces a good result. When comparing different successful extrapolations, priority is given to a good global linear fit. If none are available we pick the set with the lowest error count and take an average of both mean and error.

\begin{figure}
    \centering
    \includegraphics[scale=0.33]{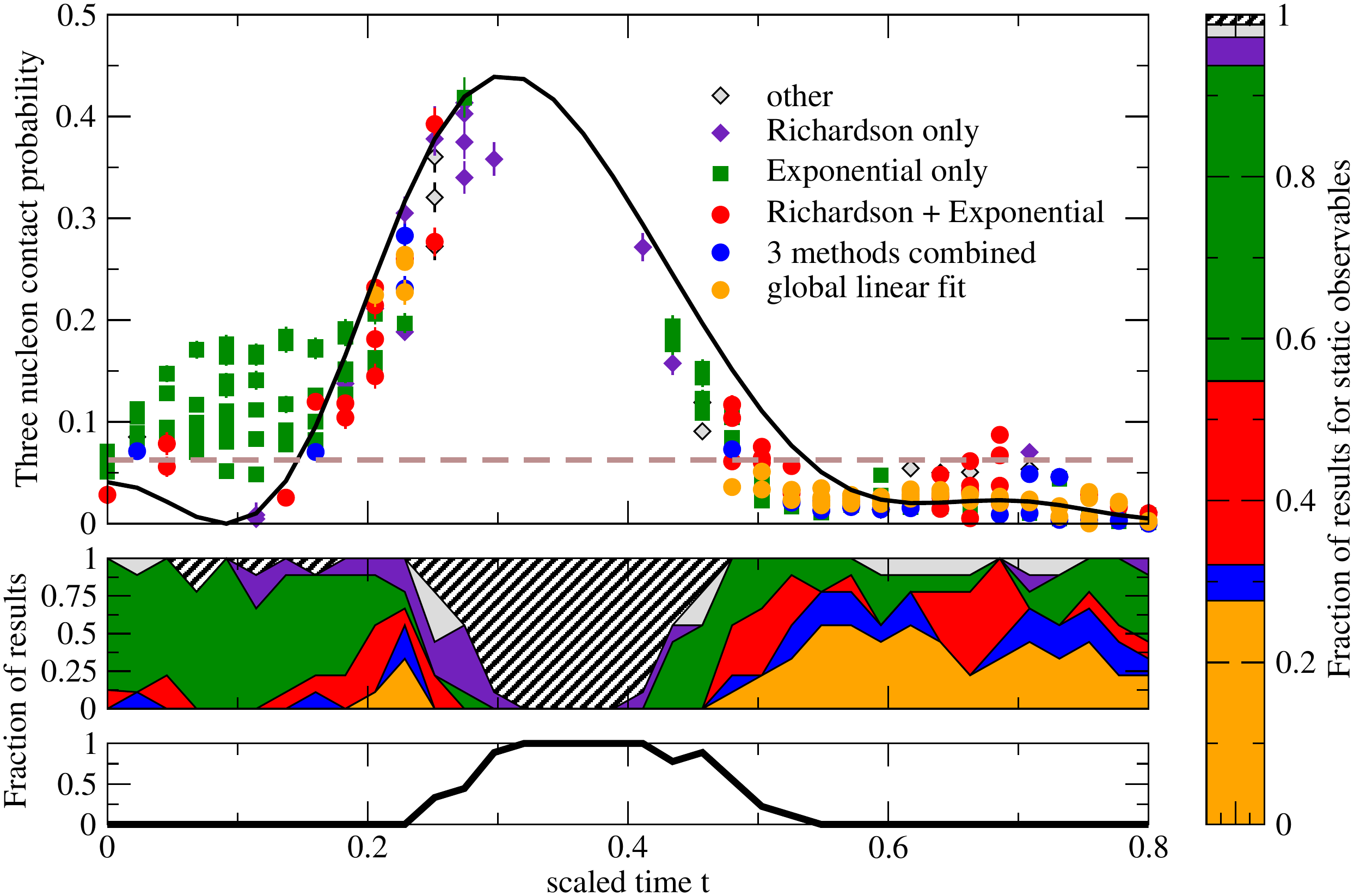}
    \caption{Extrapolation procedures used to mitigate errors in the results for the 3 nucleon contact density $C_3(t)$ shown in Fig.~\ref{fig:d0oft_cmp}. See main text for a description of the different panels. \label{fig:eextrap_cmp}}
\end{figure}

Fig.~\ref{fig:eextrap_cmp} shows the interplay between different extrapolation procedures. The main plot shows the error-mitigated probability of finding three nucleons on the same site with colors indicating the mitigation strategy employed. The central panel shows the ratios of runs mitigated with a particular strategy as a function of time. The legend of the main plot also applies here with the addition of the shaded area indicating failed runs where no stable extrapolation was possible. The bottom panel shows an estimate of the fraction of runs that have decohered (see details at the end of this section). Finally the right column shows the results for a set of $9$ static observables evaluated on the trial state showing the effect of reducing the circuit depth: for small circuits where the number of errors is not large, the small error expansion that motivates both the Richardson and Polynomial extrapolations should hold in practice. Indeed, the results for static observables show that the exponential fit is preferred on less than about $40\%$ of the calculations. When computing dynamical properties instead, the much longer circuit depth starts to favour the exponential extrapolation strategy apart from the results at late times where the magnitude of the observable is so close to the fully depolarized result that a global linear fit usually works.

In addition to the 3-body contact Eq.~\eqref{eq:3bcontact} obtained as the expectation value of the projector $P_{3B} = \rvert0000\rangle\langle0000\lvert$ shown in Fig.~\ref{fig:eextrap_cmp}, we have also computed the various 2-body contacts.
In particular we use the projector
\begin{equation}
\label{eq:p2bdyn}
\begin{split}
P_{2B-dyn} &= \rvert0101\rangle\langle0101\lvert \\
&+ \rvert1010\rangle\langle1010\lvert \\
&+ \rvert1111\rangle\langle1111\lvert\\
\end{split}
\end{equation}
to estimate the probability $C_2^D(t)$ that the two dynamical particles can be found in the same lattice site apart from the special one, and the projector
\begin{equation}
\label{eq:p2bsA}
\begin{split}
P_{2B-sA} &= \rvert0001\rangle\langle0001\lvert \\
&+ \rvert0010\rangle\langle0010\lvert \\
&+ \rvert0011\rangle\langle0011\lvert\\
\end{split}
\end{equation}
to compute the probability $C_2^A(t)$ that the first particle (tagged A here) is on the special lattice site while the other one is not (note that due to symmetry we will have the same result if we choose to tag particle B).

\begin{figure}
    \centering
    \includegraphics[scale=0.33]{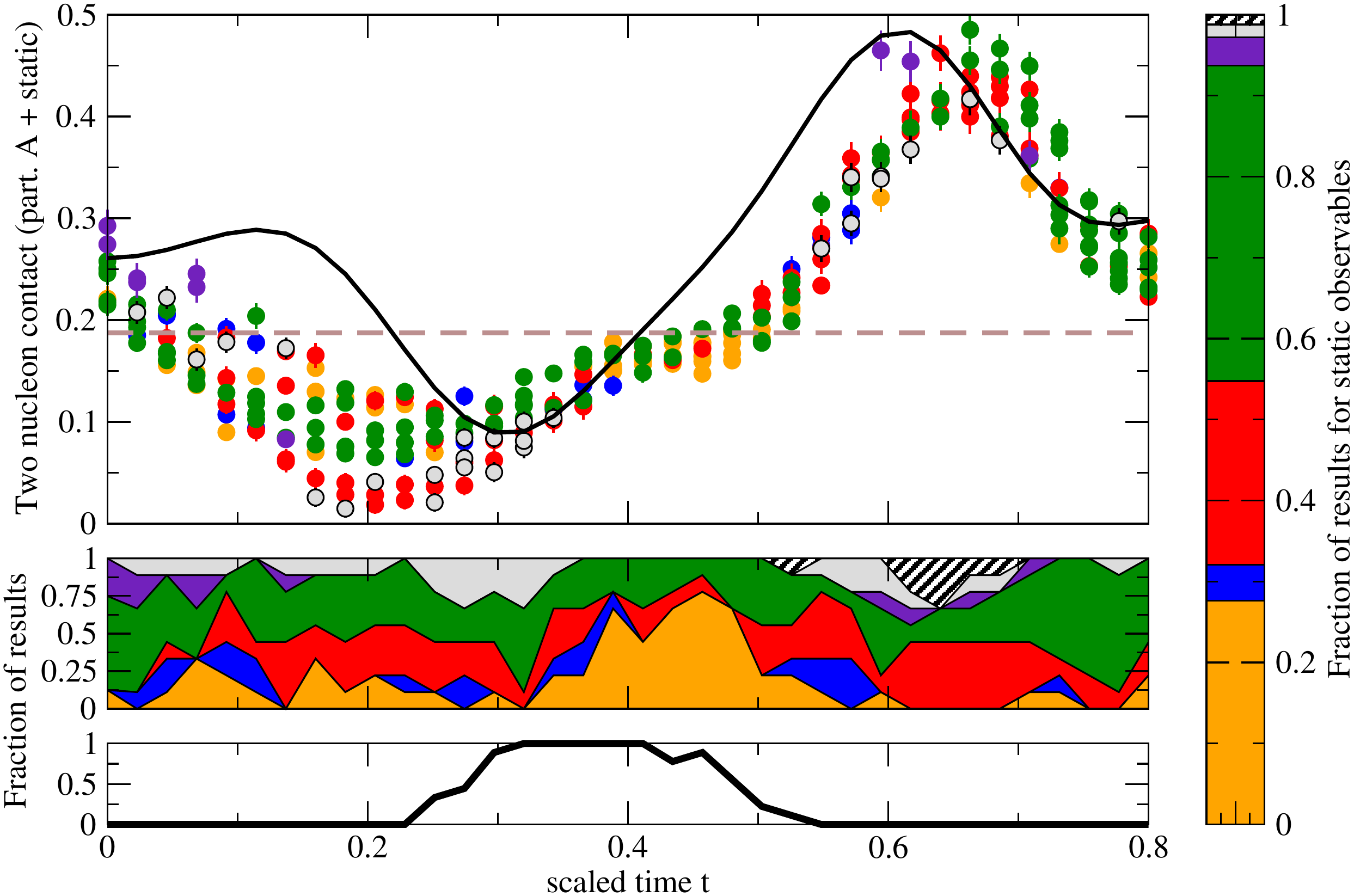}
    \caption{Extrapolation procedures used to mitigate errors in the results for the 2-body contact $P_{dyn}$ defined in Eq.~\eqref{eq:p2bdyn} of the main text. The rightmost vertical panel and the bottom horizontal panel are the same as in Fig.~\ref{fig:eextrap_cmp} and reported here for reference. \label{fig:eextrap_cmp_dyn}}
\end{figure}

\begin{figure}
    \centering
    \includegraphics[scale=0.33]{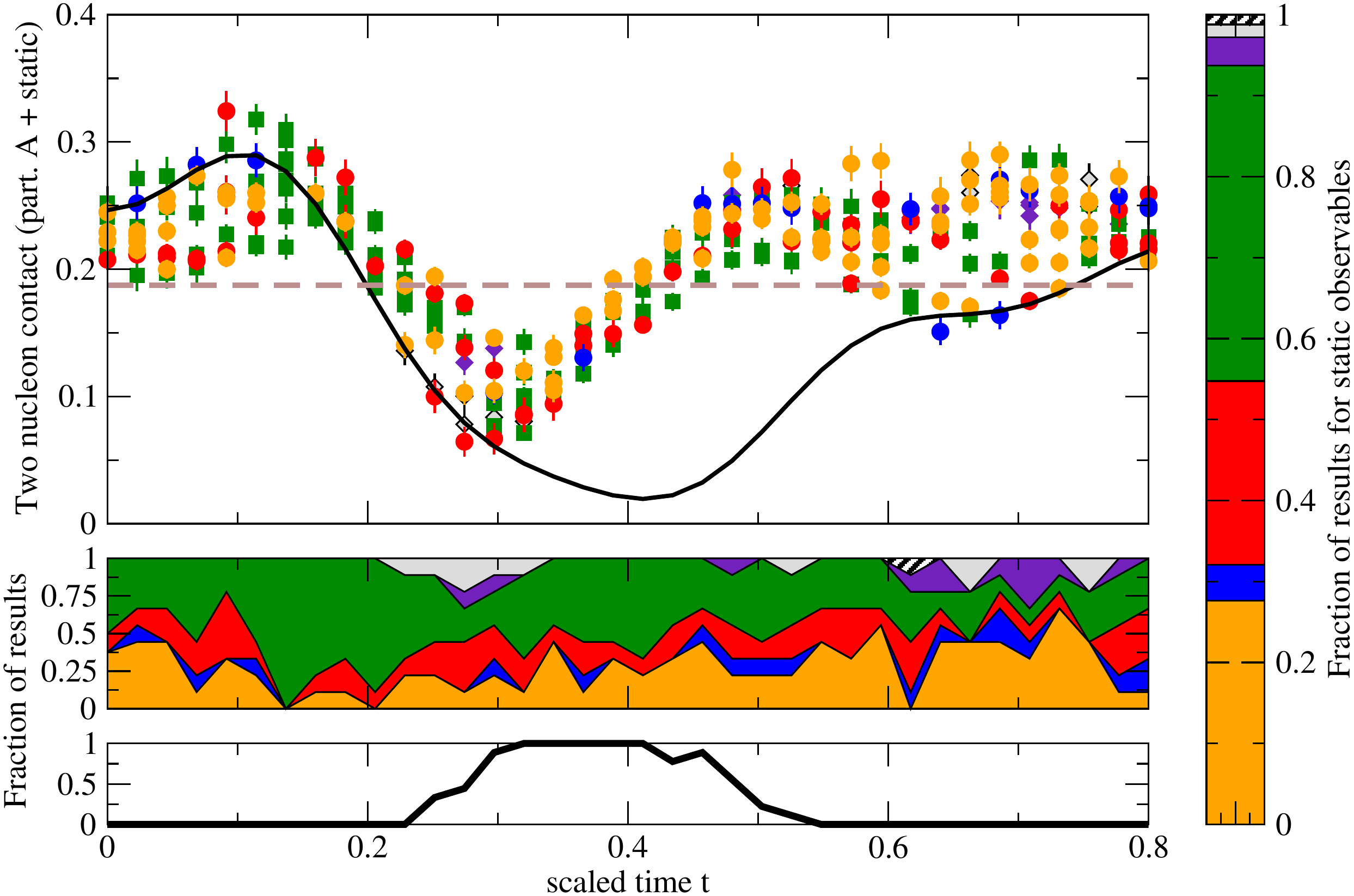}
    \caption{Extrapolation procedures used to mitigate errors in the results for the 2-body contact $P_{dyn}$ defined in Eq.~\eqref{eq:p2bsA} of the main text. The rightmost vertical panel and the bottom horizontal panel are the same as in Fig.~\ref{fig:eextrap_cmp} and reported here for reference. \label{fig:eextrap_cmp_sA}}
\end{figure}

In addition to these three extrapolations, we also check for possible complete decoherence by first checking that the distribution obtained with the smallest number of CNOT has an overlap with the fully depolarized state of less than $0.9$. We raise an error count if in the higher order results we find two distributions with overlap $>0.9$. In this work we used the trace distance as an estimator of overlap
\begin{equation}
ovd = 1 - \frac{1}{2} \sum_{i=1,16} \left\lvert \frac{1}{16} - p(i)\right\rvert
\end{equation}
with $p(i)$ are the empirical (and read-out error mitigated) probabilities. In general it might be better to include the error information in the estimator and we plan to explore different approaches in future work. For the set of runs using the mapping $[5,0,1,6]$ we found that results were possibly decohered for time in the interval $t\in[0.25,0.5]$, and this is the reason we were not able to determine robust estimators for $C_3(t)$ in that interval. This problem is not directly apparent while looking at the 2-body contact densities since the error-free result is itself close to the decohered result and therefore the test above doesn't trigger within the chosen bounds. We can, however, clearly see that the extrapolated results obtained in the problematic region are indeed compatible with the dashed brown line corresponding to the value $1/16$ as expected.

\begin{figure}
    \centering
    \includegraphics[scale=0.33]{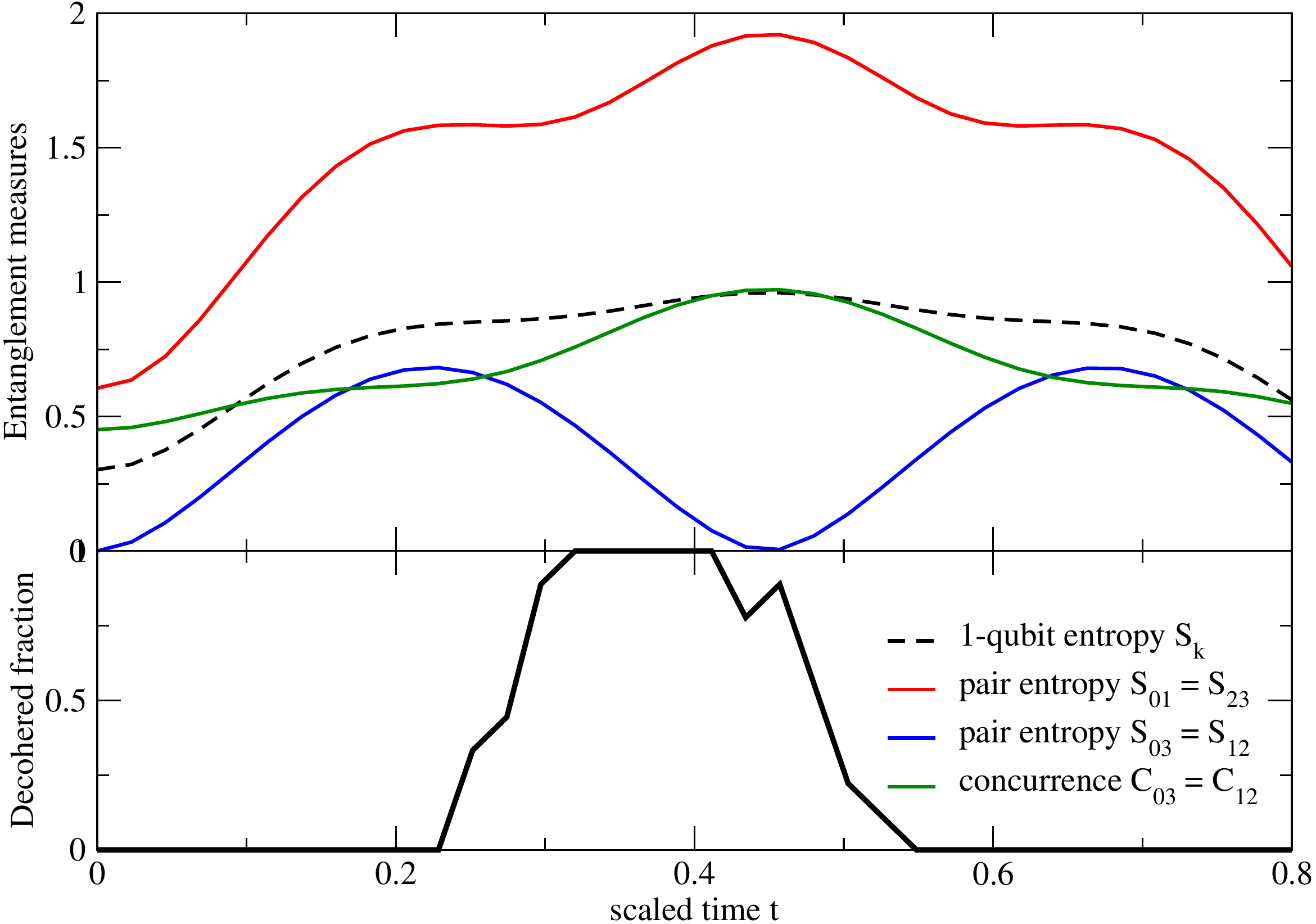}
    \caption{The top panel shows, as a function of time, different measures of entanglement obtained analytically from the expected results. The black dashed line shows the single qubit entanglement entropy, the red (blue) solid line correspond to the entanglement entropy of the pair of qubits 01 (03) and the solid green curve is the concurrence for the pair of qubits (03). For the pair 01 the concurrence is identically zero at all times. The bottom panel shows the fraction of runs where dechoerence has been detected (same as in Fig.~\ref{fig:eextrap_cmp}).\label{fig:ent}}
\end{figure}

In order to understand the systematic deviations of the experimental results form those expected in theory, we will now try to quantity the amount of entanglement generated in the time evolution. We will use different entanglement measures to study the correlations present in the 4-qubit state. The first one is the entanglement entropy which for a bipartite system described by the density matrix $\rho_{AB}$ is defined as
\begin{equation}
S_A = - \mathrm{Tr}\left[\rho_B \ln\left(\rho_B\right)\right]= -\mathrm{Tr}\left[\rho_A \ln\left(\rho_A\right)\right]=S_B
\end{equation}
where $\rho_{A\\B} = Tr_{B\\A}\left[\rho\right]$ are reduced density matrices. The entropy will be zero when the state is separable along the partition $(A,B)$. In the top panel of Fig.~\ref{fig:ent} we show the entanglement entropy for the local one-qubit density ($S_k$ with $k=\{0,1,2,3\}$) as dashed black line and the two-qubit entropies $S_{01}$ (red line) and $S_{03}$ (blue line). Note that $S_{02}=S_{01}$ due to the symmetry of trial ansatz and the Hamiltonian. We can deduce that the system starts as a product state $\rho_{in} = \rho_{03}\otimes\rho_{12}$ since $S_{03}=0$ at the start (indeed we also find that both density matrices have $rank=1$ as expected). Additionally, we see that the initial state is not extremely entangled since $S_{01}\approx 1/2$ while for a maximally entagled state it would have been 2. The time evolution initially builds up correlations between the pair of qubits $(0,3)$ and the pair $(1,2)$ as can be seen by the growing entropies along all inequivalent bipartitions and eventually leads again to a product state similar to the initial one but with much larger entanglement. Indeed around time $t\sim0.45$, both the single qubit entropies and the entropy $S_{01}$ are close to their maximum value.
To understand better these correlations we also compute the concurrence~\cite{Wooters1998} for the $3$ partitions in pairs $C_{01},C_{02},C_{03}$ (as before the first two are the same by symmetry). This measure of entanglement is defined for a 2 qubit density matrix $\rho$ as
\begin{equation}
\label{eq:concurr}
C(\rho) = \max\left\{0,\lambda_0-\lambda_1-\lambda_2-\lambda_3\right\}
\end{equation}
where $\lambda_i$ are the square root of the eigenvalues, in decreasing order, of the non-Hermitian matrix
\begin{equation}
M = \rho\left(Y\otimes Y\right)\rho^*\left(Y\otimes Y\right)
\end{equation}
and the star indicates complex conjugation. The usefulness of this measure is its relation with the entanglement of formation~\cite{Hill1997,Wooters1998} which is the minimum number of maximally-entangled pairs needed to represent $\rho$ with an ensemble of pure states~\cite{Hill1997}. In particular, the two quantities are related through the following result from Wooters~\cite{Wooters1998}:
\begin{equation}
E_F(\rho) = h\left(\frac{1+\sqrt{1-C(\rho)^2}}{2}\right)
\end{equation}
where
\begin{equation}
h(x) = -x \log_2(x) - (1-x) \log_2(1-x)
\end{equation}
and we see that $E_F$ is monotonically increasing with the concurrence $0\leq C\leq1$.

Interestingly, we find $C_{01}=C_{02}=0$ indicating that these two qubit mixed states (indeed $S_{01}$ is never zero here) do not require any entanglement to be produced (ie. the entanglement of formation is zero). The concurrence of the state $\rho_{03}$ is instead relatively large and reaches close to the value for maximally entangled states at the same position where the entropies have the maximum.

The bottom panel shows the ratio of runs that looked decohered and we can see that there is a correlation between large decoherence rate and large entanglement. A more detailed understanding of the relation between entanglement of formation and the fidelity of state preparation on non-error-corrected quantum architecture will be the subject of future work.

\section{Conclusions and Outlook}

In this paper we have attempted a first qualitative estimate of the quantum computing resources required for a minimally realistic
study of neutrino-nucleus interactions. The number of qubits required and gate counts are presented as a function of the number of nucleons and the target energy resolution in the hadronic final state. These estimates neglect completely the overhead caused by active error correction, and therefore should be considered as lower bounds on the physical resources needed for a successful execution with controllable error.

Due to the presence of substantial noise sources, full-scale studies using realistic models of nuclei and their interactions are not yet feasible with today's hardware, but showcase an important potential application of quantum computers. 

We also present results for a simple problem using present-day quantum hardware, implementing both variational algorithms for the preparation of the nuclear ground state and product formulae for the time evolution operator also required for 
calculating the response. Error mitigation strategies are presented and their beneficial impact on computation of current day machines assessed. As can be seen clearly from the results presented in Figures~\ref{fig:eextrap_cmp},~\ref{fig:eextrap_cmp_dyn} and~\ref{fig:eextrap_cmp_sA}, unitary errors can be particularly large and a more efficient strategy than what was presented here is needed to alleviate them (eg. using Pauli twirling~\cite{Wallman2016}). Another interesting outcome of our analysis of the hardware results is strong correlation between the entanglement of formation of the state being prepared and the amount of depolarizing noise that this is subject to (see. Fig.~\ref{fig:ent} and the discussion on it). This behavior is not necessarily expected for computations not using active error correction, and we plan to further explore this issue in the future.

Nevertheless, even these simple models, with simplified interactions and a small number of nucleons, allow one
to begin to understand important issues such as the importance of quantum interference in the cross sections and the expected quantum to classical transition in the examination
of explicit final states, currently handled by quasi-classical generators.
    
Further studies of the linear response and the final states are required to understand the impact of quantum computers on accelerator neutrino and related experiments. We can foresee that
quantum computers will play a significant role as their capabilities
in the number of qubits and error reduction advance.  We expect quantum linear response to be an early application of quantum computers, and neutrino scattering from nuclei to be a particularly important one.
\begin{acknowledgments}

We thank S. Pastore, A. Lovato, N. Rocco, and R. Schiavilla for discussions on neutrino-nucleus scattering, and A. Baroni for discussions on implementing nuclear models with quantum computers. In addition, we acknowledge use of the IBM Q for this work. The views expressed are those of the authors and do not reflect the official policy or position of IBM or the IBM Q team.

This research used resources of the Oak Ridge Leadership Computing Facility, which is a DOE Office of Science User Facility supported under Contract DE-AC05-00OR22725.
This project was funded in part by the DOE HEP QuantISED grant KA2401032.
This manuscript has been authored by Fermi Research Alliance, LLC under Contract No. DE-AC02-07CH11359 with the U.S. Department of Energy, Office of Science, Office of High Energy Physics.
The work of A.R. was supported by the U.S. Department of Energy, Office of Science, Office of Advanced Scientific Computing Research (ASCR) quantum algorithm teams and testbed programs, under field work proposal number ERKJ333 and by the Institute for Nuclear Theory under U.S. Department of Energy grant No. DE-FG02-00ER41132.

% There is no QuantISED number to cite according to Lali.
%\FIXME{GET DOE QUANTISED PROGRAM NUMBER CITATION RIGHT}
\end{acknowledgments}

\clearpage
%
%-----------
% reference
%-----------
\bibliography{neutrino} %%% ref.bib file
\newpage
\appendix

\section{Simple circuit primitives}
\label{app:exp_pauli}
We provide, for completeness, a few circuit representations of unitary operators we use throughout our work. The first three are exponentials of Pauli operators for which three basic gates are needed:
\begin{itemize}
\item one-body contributions
\begin{equation}
e^{-i\delta \beta Z_i}\equiv e^{-i\widetilde{\beta} Z_i} \equiv 
\Qcircuit @C=1em @R=.7em {
& \gate{\widetilde{\beta}} & \qw
}
\end{equation}
\item two body contributions
\begin{equation}
e^{-i\delta \gamma Z_iZ_j}\equiv e^{-i\widetilde{\gamma} Z_iZ_j} \equiv 
\Qcircuit @C=1em @R=.7em {
& \ctrl{1} & \qw & \ctrl{1} & \qw\\
& \targ & \gate{\widetilde{\gamma}} & \targ&\qw\\
}
\end{equation}
\item three-body contribution 
\begin{equation}
\begin{split}
e^{-i\delta t \eta Z_iZ_jZ_k}&\equiv e^{-i\widetilde{\eta} Z_iZ_jZ_k} \\
&\equiv 
\Qcircuit @C=1em @R=.7em {
& \ctrl{1} & \qw      & \qw                          & \qw      & \ctrl{1} & \qw\\
& \targ    & \ctrl{1} & \qw                          & \ctrl{1} & \targ    & \qw\\
& \qw      & \targ    & \gate{\widetilde{\eta}} & \targ    & \qw      & \qw\\
}
\end{split}
\end{equation}
\end{itemize}

Another primitive we use is the rotation around the Z axis controlled by an ancilla (see eg.~\cite{Barenco1995})
\begin{equation}
\label{eq:crotz}
\Qcircuit @C=1em @R=.7em {
& \ctrl{1} & \qw\\
& \gate{\widetilde{\gamma}} &\qw\\
}=\Qcircuit @C=1em @R=.7em {
& \qw & \ctrl{1} & \qw&\ctrl{1}&\qw\\
& \gate{\widetilde{\gamma}/2}&\targ &\gate{-\widetilde{\gamma}/2}&\targ&\qw\\
}
\end{equation}
This implementation requires 2 CNOT and 2 rotation for every controlled rotation.

\section{Details on Trotterization}
\label{app:tstep_details}
In this appendix,  we provide the detailed derivation of the Trotterization discussed in Sec.~\ref{sec:general_time_evolution}. We estimate the numbers of Trotter steps and the gate costs required for different Hamiltonian splittings and orders of the Trotter-Suzuki expansion.

To facilitate further discussion, we will consider two easy to compute upperbounds for operator norms $\|O\|$ of some hermitian operator $O=\sum_j\beta_jP_j$ written as an expansion over (tensor products of) Pauli matrices $P_j$:
\begin{itemize}
    \item absolute value norm
\begin{equation}
\left\| O \right\|_{abs}=\left\| \sum_j\beta_jP_j \right\|_{abs} = \sum_j \left\| \beta_jP_j \right\| = \sum_j \lvert \beta_j \rvert
\end{equation}
    \item physical norm
\begin{equation}
\left\| O \right\|_{phys}=sup\left\{\|O\rvert\psi\rangle\| : \ \rvert\psi\rangle \text{ physical state} \right\}
\end{equation}
where for vectors we use the 2-norm. In this context an example of a physical vector is an eigenstate of the total number operator with eigenvalue $A$.
\end{itemize}

For instance, if we consider the potential energy operator, we see that the absolute norm scales linearly with the lattice dimension
\begin{equation}
\|V\|_{abs} = M \left(6\lvert C_0\rvert+8\lvert D_0\rvert\right)\;.
\end{equation}
On the other hand, due to the fact that our interactions are contact terms, particles can interact with each other only when they are at the same lattice point and the absolute norm above will greatly overestimate the potential contribution for a physical state with $A$ fermions distributed among the $N_f=4$ types. For a physical state the maximum value of the three-body potential is reached when we occupy a lattice site with $4$ particles which means
\begin{equation}
\|V_3\|_{phys} \leq 4 \lvert D_0\rvert \left\lfloor \frac{A}{4}\right\rfloor \leq A \lvert D_0\rvert
\end{equation}
where $\lfloor x\rfloor$ is the floor function. For the two body interaction we can have either 2, 3, or 4 particles per site resulting in
\begin{equation}
\|V_2\|_{phys} \leq \lvert C_0\rvert \max \left\{\left \lfloor{\frac{A}{2}}\right \rfloor,3 \left \lfloor{\frac{A}{3}}\right \rfloor,6\left \lfloor{\frac{A}{4}}\right \rfloor\right\}\;.
\end{equation}
In the physically relevant case $M\gg A$, the physical norm will be much smaller than the absolute one. For the kinetic energy we have instead
\begin{equation}
\|K\|_{abs} = N_K \left\lvert \frac{t}{2} \right\rvert = 2DMN_f\lvert t\rvert\;,
\end{equation}
while for physical states we find instead
\begin{equation}
\|K\|_{phys} \leq A D \frac{\hbar^2}{2m}k^2_{max} = A D t \pi^2\;,
\end{equation}
where $k_{max}=\pi/a$ is the largest momentum in the box. Finally, as we did in Eq.~\eqref{eq:energy_spread}, we can use these results to place a physical upper bound on the maximum spread in energy attainable in an $A$ body system as
\begin{equation}
\begin{split}
\Delta H &= E_{max} - E_{min}\\
&= \|K\|_{phys} + \|V_2\|_{phys} + \|V_3\|_{phys} + A b_{max}\;,
\end{split}
\end{equation}
where $b_{max}$ is the nuclear binding energy at saturation and we've used the estimate $\lvert E_{min}\rvert\leq A b_{max}$ for the lowest energy value. A much better bound on the potential can be obtained by realizing that the two potential terms have opposite sign and their contributions will partially cancel, this implies that we can use
\begin{equation}
\label{eq:pot_phys_norm}
\begin{split}
\|V\|_{phys} &= \max \left\{n_2,n_3,n_4\right\}\\
&<\|V_2\|_{phys}+\|V_3\|_{phys} \;,
\end{split}
\end{equation}
where we have defined
\begin{itemize}
\item $n_2 = \lvert C_0\rvert\left \lfloor{\frac{A}{2}}\right \rfloor$
\item $n_3 = \lvert D_0 + 3C_0\rvert \left \lfloor{\frac{A}{3}}\right \rfloor$
\item $n_4 = \lvert 4D_0 + 6C_0\rvert\left \lfloor{\frac{A}{4}}\right \rfloor$
\end{itemize}
As motivated in the main text, we will use physical norms $\|\cdot\|_{phys}$ whenever possible to bound and in the following we will remove the subscript when this causes no confusion.

\subsection{Product formulae: analytical bounds}
We can now start the discussion about product formulae derived from the Trotter-Suzuki expansion for the time evolution operator. At first order, we find the simple decompositions \eqref{eq:lin_trott} and~\eqref{eq:lin_trott2} presented in the main text 
\begin{equation}
U^\alpha_L(\tau)=\prod_k^{N_K}e^{-i\tau K_k}e^{-i\tau V}\;,
\end{equation}
for the $\alpha$ splitting and 
\begin{equation}
U^\beta_L(\tau)=e^{-i\tau K}e^{-i\tau V}\;,
\end{equation}
for the $\beta$ splitting. Following the same derivation presented in~\cite{Childs2018} (see also Sec.VB of~\cite{Roggero2019} for more details), we find
\begin{equation}
\label{eq:linTS_error}
\|e^{-i\tau H} - U^{\alpha/\beta}_L(\tau) \| \leq \left(\tau \Lambda_{\alpha/\beta}\right)^2 exp\left(\tau \Lambda_{\alpha/\beta}\right)
\end{equation}
where $\Lambda_{\alpha/\beta}$ is an upper bound for the sum of norms of the individual terms in the Hamiltonian expansions. In particular, we have
\begin{align}
\Lambda_\alpha & = \sum_k\|K_k\| + \|V\| = |t|N_K +\|V\|\\
\Lambda_\beta & = \|K\| + \|V\|\;.
\end{align}

\begin{figure}
    \centering
    \includegraphics[scale=0.33]{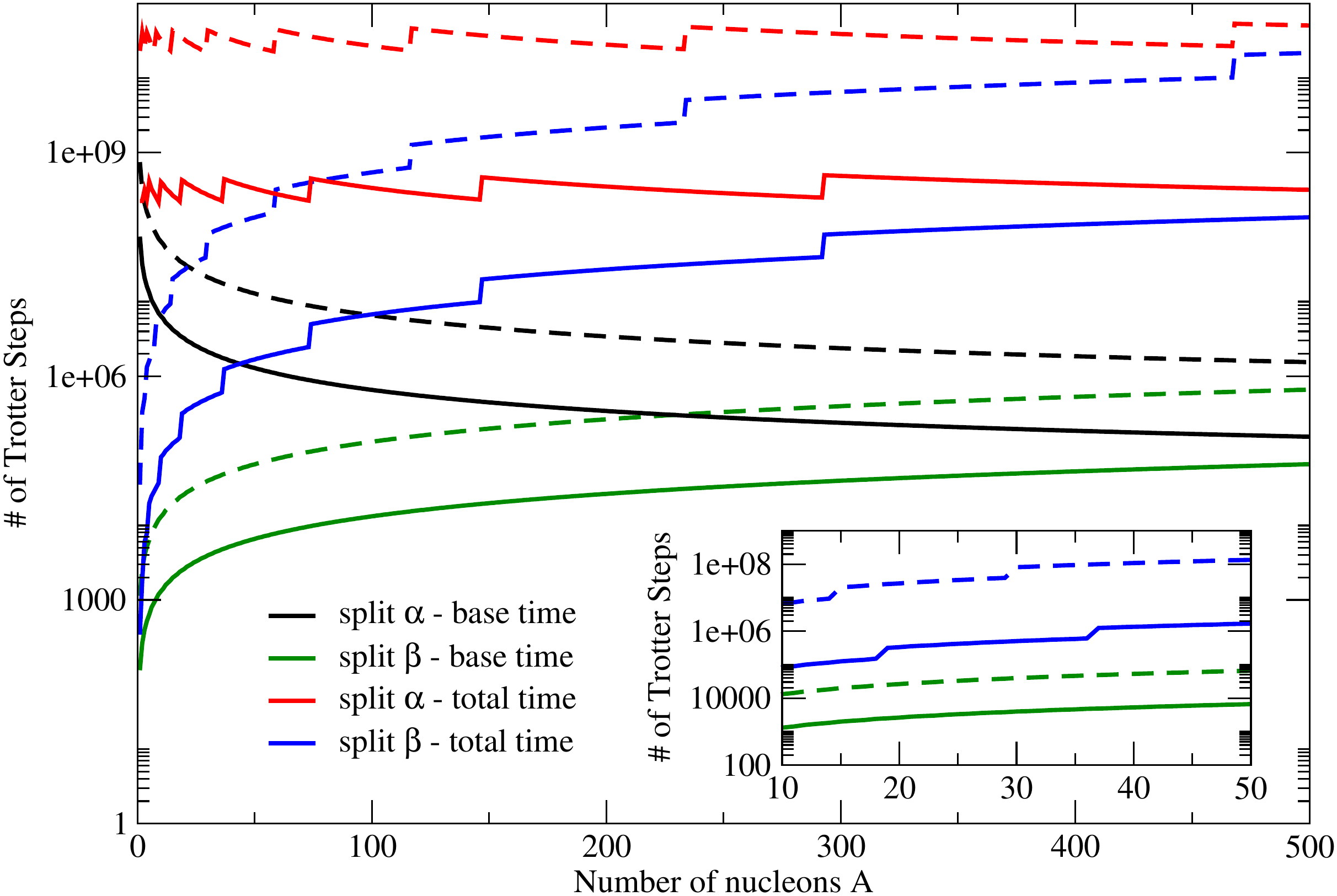}
    \caption{Estimated number of Trotter steps for both splittings of the Hamiltonian and two different target resolutions: $\delta\omega=100$ MeV (solid lines) and $\delta\omega=10$ MeV (dashed lines). The inset shows results for the best performing splitting (the $\beta$ one) in the region of mass number of interest for neutrinos.}
    \label{fig:lin_tsteps}
\end{figure}

At this point, we first note that we can interpret the evolution under the approximate propagator $U_L(\tau)$ as an exact time evolution under the effective Hamiltonian (cf.~\cite{Wecker2014}) given by
\begin{equation}
H^{\alpha/\beta}_{\mathrm{eff}}(L) = \frac{\ln\left(U^{\alpha/\beta}_L(\tau)\right)}{-i\tau}
\end{equation}
and for small values of $\tau$, we estimate the error in the energy eigenvalues using (see also~\cite{kivlichan2019improved} for a tighter bound)% was \cite{1902.10673}
\begin{equation}
\|H - H^{\alpha/\beta}_{\mathrm{eff}}(L)\| = \frac{1}{\tau}\|e^{-i\tau H} - U^{\alpha/\beta}_L(\tau) \|\;.
\end{equation}
In order to control the approximation error introduced by using the approximate evolution operator $U_L$, we can split the time interval $\tau$ required by our algorithm into $r$ steps and consider instead
\begin{equation}
\|e^{-i\tau H} - U^{\alpha/\beta}_L(\tau/r)^r \| =\delta_\tau \;,%\leq \frac{\left(\tau \Lambda_{\alpha/\beta}\right)^2}{r} exp\left(\frac{\tau \Lambda_{\alpha/\beta}}{r}\right)\;,
\end{equation}
leading to an energy error $\epsilon_\tau$ bounded by
\begin{equation}
\frac{\delta_\tau}{\tau} \leq \frac{\tau}{r}\Lambda_{\alpha/\beta}^2 \exp\left(\frac{\tau \Lambda_{\alpha/\beta}}{r}\right)\;.
\end{equation}
Following the same analysis presented in~\cite{Childs2018}, we obtain the following analytical bound for the number of Trotter steps needed for time $\tau$ 
\begin{equation}
\label{eq:an_bnd_lin}
r_{1;A} = \left\lceil \max\left\{\tau \Lambda_{\alpha/\beta},\frac{e\tau \Lambda_{\alpha/\beta}^2}{\epsilon_\tau}\right\}\right\rceil\;.
\end{equation}
In Fig.~\ref{fig:lin_tsteps}. we present the expected number of steps needed to perform time evolution for both the base time $\tau_{base}=2\pi/\Delta H$ (black and green lines) and the whole sequence of $W$ evolutions for a total time of $\tau_{tot}=(2^W-1)*\tau_{base}$ (red and blue lines) where the number of ancilla qubits $W$ is obtained for a fixed resolution $\Delta\omega$ (cf. Eq.~\eqref{eq:num_anc} in the main text). Results are presented for the hardest interaction ($a=1.4$ fm) and for two different target resolutions: $\delta\omega=100$ MeV (solid lines) and $\Delta\omega=10$ MeV (dashed lines). In both cases, we fix the energy error $\epsilon_\tau$ to be half the resolution.
% TODO? - I think this is okay...
%\ale{add some comments to the plot, if needed have a plot of $W(A)$}

%\FIXME{One can also define an empirical bound
%\begin{equation}
%r_{1;E} = \min\left\{r\in\mathbb{N}:\frac{\tau}{r}\Lambda^2 exp\left(\frac{\tau \Lambda}{r}\right)\leq\epsilon_\tau\right\}\;.
%\end{equation}
%This bound is not much better than the analytical one in many cases, we won't consider it any longer here.}

Higher order decompositions allow for a reduction of the scaling of the approximation error with the evolution time and can therefore provide an important efficiency gain. Here we will consider the even order Trotter-Suzuki formulae~\cite{Suzuki91} defined by the recursion
\begin{equation}
\label{eq:symm_prod_form}
S_{2k}(\tau) = \left[S_{2k-2}(\tau_k)\right]^2S_{2k-2}\left(\tau-4\tau_k\right)\left[S_{2k-2}(\tau_k)\right]^2
\end{equation}
with
\begin{equation}
S_2(\tau) = \prod_{k=1}^{N_K}e^{-i\frac{\tau}{2} K_k}e^{-i\frac{\tau}{2} V}e^{-i\frac{\tau}{2} V}\prod_{k=N_K}^{1}e^{-i\frac{\tau}{2} K_k}    
\end{equation}
and $\tau_k=\tau/(4-4^{1/(2k-1)})$ for $k>1$. Using these approximations (cf.~\cite{Childs2018,Roggero2019}) to the evolution operator the number of steps needed for a given accuracy becomes bounded by
\begin{equation}
\label{eq:an_bnd_quad}
r_{2k;A} = \left\lceil \rho_{2k}\max\left\{1,\left(\frac{2e\Lambda 5^{k-1}}{3\epsilon_\tau}\right)^{\frac{1}{2k}}\right\}\right\rceil
\end{equation}
with $\rho_{2k}=2\tau5^{k-1} \Lambda$.%, or by the possible tighter bound
%\begin{equation}
%r_{2k;E} = \min\left\{r\in\mathbb{N}:\frac{\rho_{2k}^{2k+1}}{3r^{2k}}\Lambda^2 %exp\left(\frac{\rho_{2k}}{r}\right)\leq\epsilon_\tau  \right\}\;.
%\end{equation}

As reported in the main text, the explicit expressions for the second order formulae with both kinds of breakup are
\begin{equation}
S_\alpha(\tau) = e^{-i\frac{\tau}{2} V}\prod_{k=1}^{N_K}e^{-i\frac{\tau}{2} K_k}\prod_{k=N_K}^{1}e^{-i\frac{\tau}{2} K_k}e^{-i\frac{\tau}{2} V}
\end{equation}
for the $\alpha$ splitting, while for the $\beta$ splitting we consider the two options
\begin{equation}
S^{K+V}_\beta(\tau) = e^{-i\frac{\tau}{2} K}e^{-i\tau V}e^{-i\frac{\tau}{2} K}
\end{equation}
and
\begin{equation}
S^{V+K}_\beta(\tau) = e^{-i\frac{\tau}{2} V}e^{-i\tau K}e^{-i\frac{\tau}{2} V}\;.
\end{equation}
These formulae are used to produce the results reported next.

\begin{figure}
    \centering
    \includegraphics[scale=0.33]{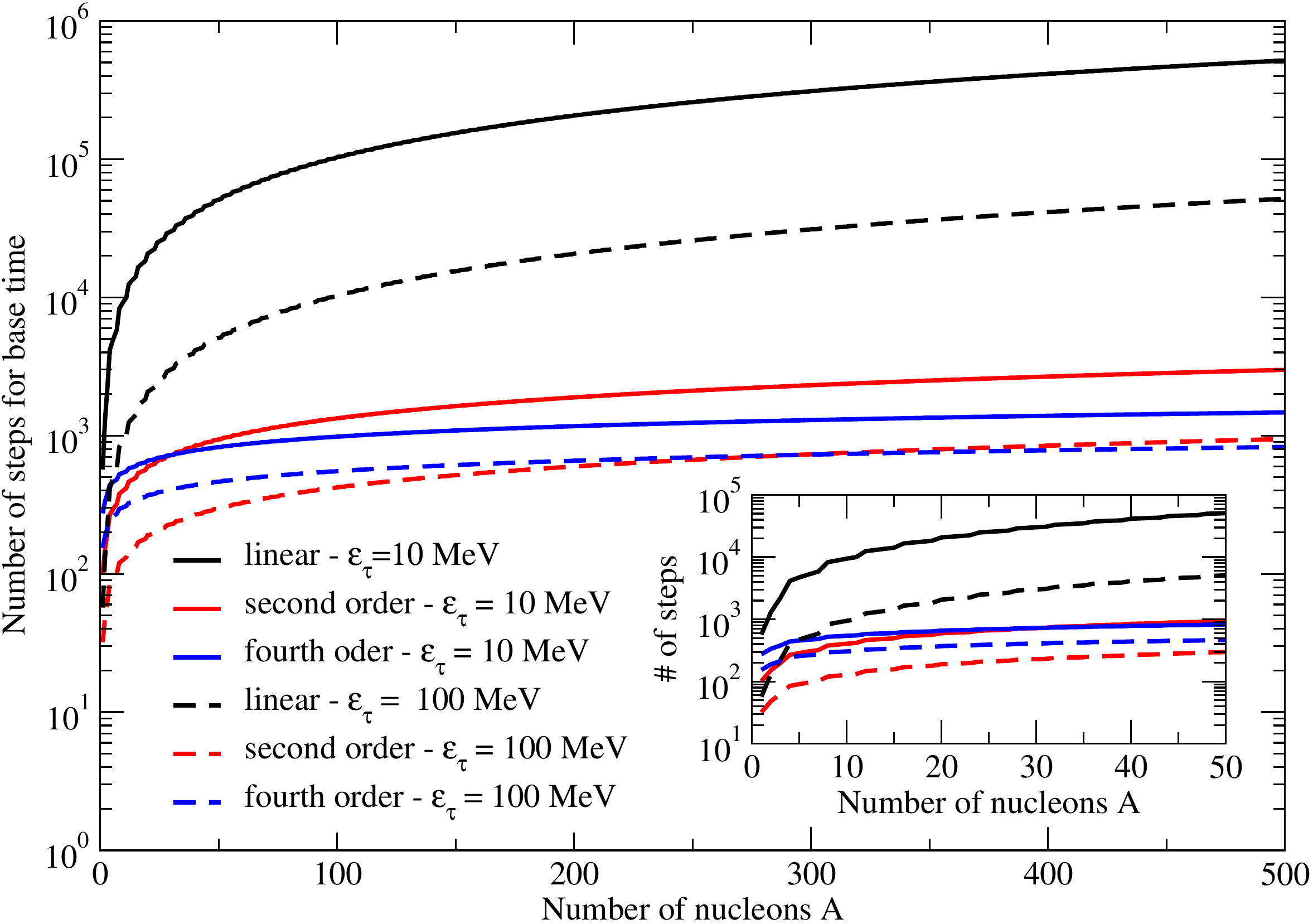}
    \caption{Number of Trotter steps required for a target precision $\epsilon_\tau=10$ MeV ($100$ MeV) using $\beta$ splitting shown as solid (dashed) lines for the base time $\tau=2\pi/\Delta H$. }
    \label{fig:A}
\end{figure}

In Fig.~\ref{fig:A}, we show results for $r_{1;A}$, $r_{2;A}$ and $r_{4;A}$ for the base time interval needed for $\beta$ splitting with $\tau = 2 \pi / \Delta H$ using the hardest interaction with $a=1.4$ fm (softer interactions require a smaller number of steps due to the smaller norm of the Hamiltonian).
% \FIXME{Some more comments on plot}
The second order formulas (red lines) shows a clear advantage over the linear decompositions (black lines). On the other hand, the fourth order formulas (blue lines) becomes favorable only when tackling big enough problems. Specifically, the break-even point is $A=24$ ($A=234$) for higher target accuracy $\delta\omega=10$ MeV (lower target accuracy $\delta\omega=100$ MeV).

\begin{figure}
    \centering
    \includegraphics[scale=0.33]{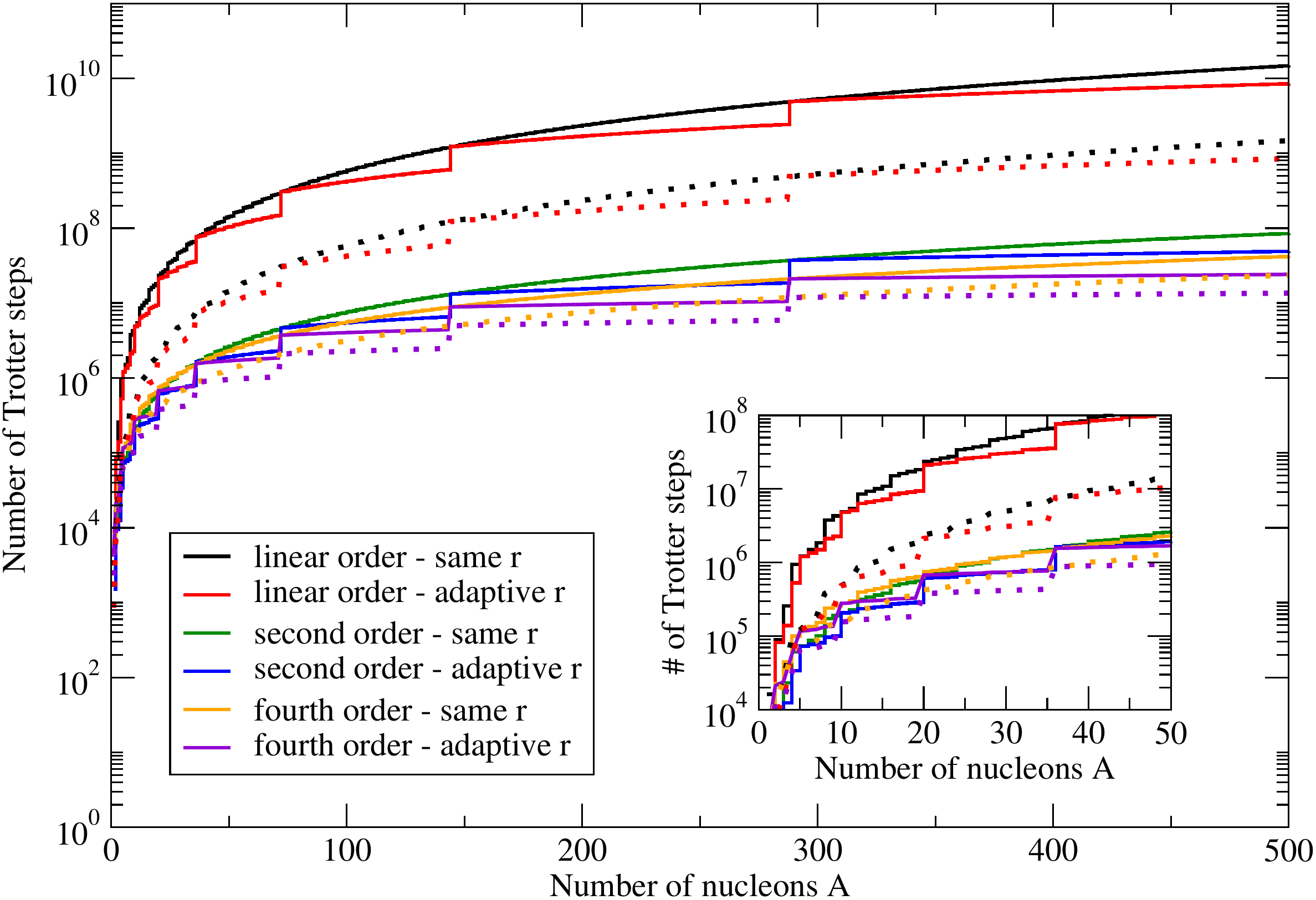}
    \caption{Total number of Trotter steps required for a target precision $\epsilon_\tau=10$ MeV ($100$ MeV) shown as solid (dotted) lines. In contrast to Fig~\ref{fig:A} these results are obtained for the full propagation time $T=2^{W-1}\tau$. See text for details.
    %	\FIXME{double check these plots}
    }
    \label{fig:B}
\end{figure}

During the Phase Estimation stage of our algorithm, we need to perform (controlled) time evolution for a set of $N_{tot} = 2^{W-1} = \Delta H / \epsilon_\tau$ (note the factor of two coming from $\epsilon_\tau=\Delta\omega/2$) time intervals given by $T_k=2^k\tau$ for $k\in[0,W-1]$. One way to achieve this is to decompose optimally the unitary operator $U_L(\tau)$ using the bounds presented above and simply repeat this basic one as needed. The resulting total number of steps required by this algorithm is denoted as "same r" in Fig.~\ref{fig:B}. An alternative approach is to adaptively decompose each of the $N_{tot}$ evolution unitary operators individually and then sum the number of steps together. This method produces the results denoted as "adaptive r" in Fig.~\ref{fig:B} and, as expected, is usually more efficient than the simpler standard one. This is the strategy used throughout the main text.

\subsection{Product formulae: commutator bounds}
As mentioned in Sec.~\ref{sec:general_time_evolution}, the errors in product formulas should depend on the commutators of the terms in the Hamiltonian and not directly on their norms, this is one of the deficiencies of the bounds considered above and prevents them from being tight. For the linear Trotter decomposition, we can consider the commutator bound similar to the one from~\cite{Childs2018}:
\begin{equation}
\label{eq:linTS_error_comm}
\begin{split}
\|e^{-i\tau H} - &U^{\alpha/\beta}_L(\tau) \| \leq\\
&\frac{C_{\alpha/\beta}}{2}\tau^2+\frac{\left(\tau \Lambda_{\alpha/\beta}\right)^3}{3} exp\left(\tau \Lambda_{\alpha/\beta}\right)
\end{split}
\end{equation}
where
\begin{equation}
\begin{split}
C_\alpha &= \|\sum_{j>k}\left[H_k,H_j\right]\| = \|\sum_{k}^{N_K}\left[K_k,V\right]+\sum_{j>k}\left[K_k,K_j\right]\|\\
&\leq 2N_K \lvert t \rvert\|V\|_{phys} + N_K(N_K-1) t^2 \leq \Lambda^2_\alpha
\end{split}
\end{equation}
for the $\alpha$ splitting, and for the $\beta$ splitting
\begin{equation}
C_\beta = \|\left[K,V\right]\| \leq 2\|K\|_{phys}\|V\|_{phys} \leq \Lambda^2_\beta\;.
\end{equation}
We can now estimate the number of intervals $r$ by defining
\begin{equation}
r_{1;C} = \min\left\{r\in\mathbb{N}: \Gamma^1_{\alpha/\beta}(r)\leq\epsilon_\tau\right\}\;,
\end{equation}
where $\Gamma^1_{\alpha/\beta}$ is the error estimator obtained from the upperbound Eq.~\eqref{eq:linTS_error_comm} above:
\begin{equation}
\Gamma^1_{\alpha/\beta}(r) = \frac{C_{\alpha/\beta}}{2}\frac{\tau}{r}+\left(\frac{\tau}{r}\right)^2\frac{\left( \Lambda_{\alpha/\beta}\right)^3}{3} exp\left(\frac{\tau}{r} \Lambda_{\alpha/\beta}\right)\;,
\end{equation}
and for $C_{\alpha/\beta}$ we use their upperbounds derived above.

\begin{figure}
    \centering
    \includegraphics[scale=0.33]{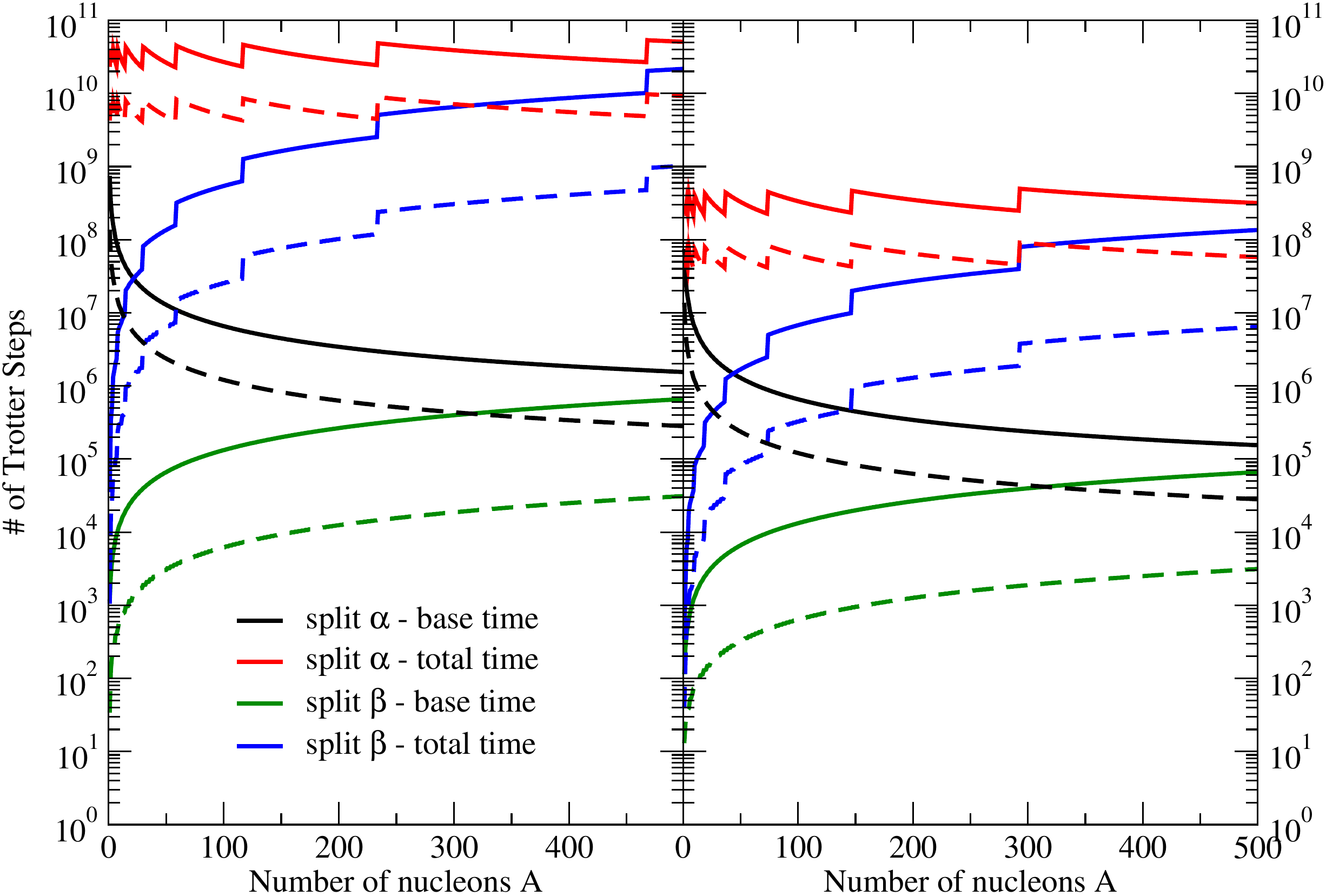}
    \caption{Comparison of the analytical vs. commutator bounds $r_{1:A}$ and $r_{1:C}$  for the linear Trotter-Suzuki breakup with both splitting schemes. The left panel corresponds to a target resolution $\Delta\omega=10$ MeV while the right panel to $\Delta\omega=100$ MeV.}
    \label{fig:lin_comm_tsteps}
\end{figure}

The importance of including information about the commutators is apparent from the results in Fig.~\ref{fig:lin_comm_tsteps} where we show the improved bounds $r_{1;C}$ (dashed lines) together with the analytical results $r_{1;A}$ (solid lines) for the two target precisions separately (the left panel correspond to $\Delta\omega=10$ MeV while the right panel to $\Delta\omega=100$ MeV). The adoption of the commutator bounds provides an improvement of the same order of magnitude as going to a second order expansion (cf. Fig.~\ref{fig:quad_tsteps} in the main text).

For the second order expansion instead we use the result of \cite{kivlichan2019improved} to construct the estimators
\begin{equation}
r_{2;C} = \min\left\{r\in\mathbb{N}: \Gamma^2_{\alpha/\beta}(r)\leq\epsilon_\tau\right\}\;,
\end{equation}
where the error estimators are given by
\begin{equation}
\Gamma^2_{\alpha/\beta}(r) = \frac{1}{12}\left(\frac{\tau}{r}\right)^2 T_{\alpha/\beta}\;,
\end{equation}
with
\begin{equation}
\begin{split}
T_\alpha &= \sum_{j,k}^{N_K}\|\left[\left[V,K_j\right],K_k\right]\| + \sum_{j}^{N_K}\|\left[\left[V,K_j\right],V\right]\| \\
&+ \sum_{i}^{N_K} \sum_{j>i}\sum_{k>i}\|\left[\left[K_i,K_j\right],K_k\right]\|\\
&+ \sum_{i}^{N_K}\sum_{j>i}^{N_K}\|\left[\left[K_i,K_j\right],K_i\right]\|\\
&\leq 4N_K^2 t^2 \|V\|_{phys} + 4N_K \lvert t\rvert \|V\|^2_{phys} +2N_K(N_K-1) \lvert t\rvert^3\\
&+\frac{2}{3}N_K\left(2N_K^2-3N_K+1\right)\lvert t\rvert^3
\end{split}
\end{equation}
and two different expressions for splitting $\beta$ depending on the ordering of the operators
\begin{equation}
\begin{split}
T^{K+V}_\beta &= \|\left[\left[K,V\right],V\right]\| + \frac{1}{2}\|\left[\left[K,V\right],K\right]\| \\
&\leq 2\|K\|_{phys}\|V\|_{phys} \left(2\|V\|_{phys}+\|K\|_{phys}\right)
\end{split}
\end{equation}
and
\begin{equation}
\begin{split}
T^{V+K}_\beta &= \|\left[\left[V,K\right],K\right]\| + \frac{1}{2}\|\left[\left[V,K\right],V\right]\| \\
&\leq 2\|K\|_{phys}\|V\|_{phys} \left(2\|K\|_{phys}+\|V\|_{phys}\right)\;.
\end{split}
\end{equation}
We show the resulting estimates for $r_{2;C}$ at a fixed target accuracy $\Delta\omega=100$ MeV for the two splitting methods in Fig.~\ref{fig:second_comm_tsteps} on the main text.

\subsection{Product formulas: gate cost per step}
\label{app:tstep_impl}
In order to implement the time-evolution unitary operators described in the preceding section, we need to implement three independent unitary operators
\begin{align}
U_1(\tau)&=e^{-i\tau V}\\
U_2(\tau)&=e^{-i\tau K}\\
U_3(\tau)&=\prod_{k=1}^{N_K}e^{-i\tau K_k}\\
U_4(\tau)&=\prod_{k=N_K}^1e^{-i\tau K_k}=U^\dagger_3(-\tau) \,,
\end{align}
from which we construct
\begin{align}
U^\alpha_L(\tau)&=U_3(\tau)U_1(\tau)\\
U^\beta_L(\tau)&=U_2(\tau)U_1(\tau)\\
S_\alpha(\tau)&=U_1\left(\frac{\tau}{2}\right)U_3\left(\frac{\tau}{2}\right)U_4\left(\frac{\tau}{2}\right)U_1\left(\frac{\tau}{2}\right)\\
S^{K+V}_\beta(\tau)&=U_2\left(\frac{\tau}{2}\right)U_1(\tau)U_2\left(\frac{\tau}{2}\right)\\
S^{V+K}_\beta(\tau)&=U_1\left(\frac{\tau}{2}\right)U_2(\tau)U_1\left(\frac{\tau}{2}\right)\;.
\end{align}

\subsubsection{Evolution operator for the interaction term}
\label{app:u1imp}
Since the interactions in our model have zero range, all the $M$ distinct potential energy operators acting on different sites will commute:
\begin{equation}
U_1(\tau)=e^{-i\tau V}=\prod_{i=1}^M e^{-i\tau V_i}
\end{equation}
with $V_i$ acting non-trivially on only $N_f$ qubits at a time. This implies that we need to worry about the implementation of only a single diagonal Hamiltonian term of the following form
\begin{equation}
\label{eq:pot_decomp}
\begin{split}
V_i &= \alpha + \beta \sum_{f=0}^3 Z_f + \gamma \sum_{f>f'}^3 Z_f Z_{f'}\\
&+ \eta \sum_{f>f'>f''}^3 Z_f Z_{f'}Z_{f''} \;,
\end{split}
\end{equation}
where the coefficient can be read directly from the general expression in Eq.~\eqref{eq:pot_energy}. Using well known general decompositions for exponentials of Pauli matrices discussed in Appendix~\ref{app:exp_pauli},
% (see eg.~\cite{Somma2002}
we can express the evolution operator $U_i(\delta)=\exp\left(-i\delta V_i\right)$ in terms of single qubit Z-rotations and CNOT gates. Assuming all to all connectivity within the 4 qubit cell (and with the possible controlling ancilla), a straightforward implementation using the above mentioned gadgets will require 14 rotations and 28 CNOT for the uncontrolled version and 28 rotations and 56 CNOT for the controlled unitary evolution. We know however that the optimal circuit to implement an arbitrary diagonal unitary on $n$ qubits requires at most $2^{n+1}-3$ one and two qubit gates (see~\cite{Bullock2004,Welch2014}). Given the lack of a four-body contact interaction, and assuming all-to-all connectivity, for our model this expansion produces the following circuit with depth 28 (14 rotations + 14 CNOT) 
\begin{widetext}
\Qcircuit @C=0.8em @R=.7em {
&\gate{\beta}&\ctrl{1}& \qw &\ctrl{1}&\qw  &\qw&\qw&\ctrl{2}&\qw&\qw&\qw&\ctrl{2}&\qw&\qw&\qw&\ctrl{3}&\qw&\qw&\ctrl{3}&\qw&\qw&\qw&\ctrl{3}&\qw&\qw&\qw&\ctrl{3}&\qw&\qw\\
&\qw&\targ&\gate{\gamma}&\targ&\gate{\beta}&\ctrl{1}&\qw&\qw&\qw&\ctrl{1}&\qw&\qw&\qw&\qw&\qw&\qw&\qw&\ctrl{2}&\qw&\qw&\qw&\qw&\qw&\qw&\ctrl{2}&\qw&\qw&\qw&\qw\\
&\qw &\qw&\qw&\qw&\qw&\targ&\gate{\gamma}&\targ&\gate{\eta}&\targ&\gate{\gamma}&\targ&\gate{\beta}&\ctrl{1}&\qw&\qw&\qw&\qw&\qw&\qw&\ctrl{1}&\qw&\qw&\qw&\qw&\qw&\qw&\qw&\qw\\
&\qw&\qw&\qw&\qw&\qw&\qw&\qw&\qw&\qw&\qw&\qw&\qw&\qw&\targ&\gate{\gamma}&\targ&\gate{\eta}&\targ&\targ&\gate{\eta}&\targ&\gate{\gamma}&\targ&\gate{\eta}&\targ&\gate{\gamma}&\targ&\gate{\beta}&\qw
}
\end{widetext}
where the one qubit gates are appropriate Z rotations (cf. Eq.~\eqref{eq:pot_decomp}). Under the more stringent constraint of 2D planar connectivity one can optimize the construction for parallel efficiency, the result of this exercise (first reported in~\cite{Schuch2002}) is the following circuit
\begin{widetext}
\begin{equation}
\label{eq:optimal_pot_circuit}
\Qcircuit @C=0.8em @R=.7em {
&\ctrl{1}&\gate{\beta}&\targ&\qw&\gate{\eta}&\ctrl{1}&\qw&\targ&\qw&\qw&\ctrl{1}&\gate{\gamma}&\targ&\qw&\ctrl{1}&\gate{\gamma}&\targ&\qw&\qw&\qw\\
&\targ&\gate{\gamma}&\qw&\targ&\gate{\eta}&\targ&\gate{\gamma}&\qw&\targ&\gate{\eta}&\targ&\gate{\eta}&\qw&\targ&\targ&\gate{\gamma}&\qw&\targ&\gate{\beta}&\qw\\
&\ctrl{1}&\qw&\qw&\ctrl{-1}&\gate{\beta}&\ctrl{1}&\qw&\qw&\ctrl{-1}&\qw&\ctrl{1}&\qw&\qw&\ctrl{-1}&\ctrl{1}&\qw&\qw&\ctrl{-1}&\qw&\qw\\
&\targ&\qw&\ctrl{-3}&\qw&\gate{\gamma}&\targ&\qw&\ctrl{-3}&\qw&\qw&\targ&\qw&\ctrl{-3}&\qw&\targ&\gate{\beta}&\ctrl{-3}&\qw&\qw&\qw
}
\end{equation}
\end{widetext}
which has serial (parallel) depth of 30 (15) with a 2D nearest neighbor connectivity (7 rotations + 8 CNOT in parallel).
The gate cost of implementing the potential energy propagator is summarized in Tab.~\ref{table:U1_gate_count}.

\begin{table}[]
\begin{tabular}{|l|cc|cc|}
\hline
 & \multicolumn{2}{c|}{\# c-Rz} & \multicolumn{2}{c|}{\# CNOT} \\
 &    serial       &    parallel       &    serial       &   parallel        \\ \hline
naive &     14M      &     14      &   28M        &   28       \\
serial opt.   &     14M      &     14      &  14M         &   14       \\
parallel opt. &     14M      &      7     &   16M       &      8 \\\hline

\end{tabular}
\caption{Gate cost for the potential energy propagator.}
\label{table:U1_gate_count}
\end{table}

\subsubsection{Evolution operator for the hopping term}
We will, for now, only assume linear connectivity for our implementation of the last $3$ evolution operators.
%\FIXME{There are known algorithms with O(1) depth with planar connectivity for a 2D geometry by doubling the number of qubits, I don't know how much of this can be ported to 3D.}
The exact propagator for hopping term $U_2(\tau)$ can be obtained using either the FFFT whenever $4M$ is power of $2$ or else the Givens rotations described in \cite{Kivlichan2018,kivlichan2019improved}. In the latter case, due to our choice of ordering where single particle states on the same lattice point and different spin-isospin are next to each other (this is for ease of implementation of the potential energy part, especially the triples), we consider the system as $4M$ spinless fermions and use all the $2M(4M-1)$ Givens rotations each requiring $2$ rotations (which can be performed in parallel) and 5 CNOT for a total of $4M(4M-1)$ arbitrary Z-rotations and $10M(4M-1)$ CNOT with parallel depth $8M-3$ (using results from~\cite{Kivlichan2018}). This circuit must be executed twice and a final set of $4M$ rotations in depth $1$ must be performed in between. Luckily, only these need to be controlled when performing the controlled time-evolution. The gate cost of implementing the propagator $U_2$ is summarized in Tab.~\ref{table:U2_gate_count}.

\begin{table}[]
	\begin{tabular}{|l|cc|cc|cc|}
	\hline
	& \multicolumn{2}{|c|}{\# Rz}& \multicolumn{2}{c|}{\# c-Rz} & \multicolumn{2}{c|}{\# CNOT} \\
	& serial       &    parallel      &    serial       &    parallel       &    serial       &   parallel        \\ \hline
	$U_2$ &	32$M^2$      &     16M-4  & 0 & 0 & 20M(4M-1)        &   80M-28   \\ \hline
	c-$U_2$&	8M(4M-1)      &     16M-6 &     4M      &     1      &   4M(20M-3)   &   80M-30 \\ \hline
	\end{tabular}
	\caption{Gate cost for the $U_2$ propagator.}
	\label{table:U2_gate_count}
\end{table}

%\begin{table}[]
%	\begin{tabular}{|cc|cc|}
%		\hline
%		\multicolumn{2}{|c|}{\# Rz} & \multicolumn{2}{c|}{\# CNOT} \\
%		serial       &    parallel    &    serial       &   parallel        \\ \hline
%		32$M^2$      &     16M-4  &   20M(4M-1)        &   80M-28   \\ \hline
%	\end{tabular}
%	\caption{Gate cost for the $U_2$ propagator.}
%\end{table}
%
%\begin{table}[]
%\begin{tabular}{|cc|cc|cc|}
%\hline
% \multicolumn{2}{|c|}{\# Rz}& \multicolumn{2}{c|}{\# c-Rz} & \multicolumn{2}{c|}{\# CNOT} \\
%   serial       &    parallel      &    serial       &    parallel       &    serial       &   parallel        \\ \hline
%    8M(4M-1)      &     16M-6 &     4M      &     1      &   4M(20M-3)   &   80M-30 \\
%    \hline      
%\end{tabular}
%\caption{Gate cost for the $U_2$ propagator.}
%\end{table}

We turn now to the implementation of $U_3$, a naive implementation of all the $N_K$ terms separately would require $2$ rotations (by the same angle), 8 Hadamard and 4 S gates and $4(p_k-q_k)$ CNOT for each term 
\begin{equation}
e^{-i\tau K_k} = e^{-i\frac{\tau}{2} (X_{p_k}X_{q_k}+Y_{p_k}Y_{q_k}) Z_{p_k+1}\cdots Z_{q_k-1}}
\end{equation}
in the expansion of $U_3(\tau)$, where the string of Pauli Z comes from the Jordan-Wigner mapping. This estimate comes from the following explicit construction (cf.~\cite{Wecker2014})
\begin{widetext}
\begin{equation}
\Qcircuit @C=0.8em @R=.7em {
&\gate{H}&\ctrl{1}&\qw&\qw&\qw&\qw&\qw&\qw&\qw&\ctrl{1}&\gate{HSH}&\ctrl{1}&\qw&\qw&\qw&\qw&\qw&\qw&\qw&\ctrl{1}&\gate{HS^\dagger}&\qw\\
&\qw&\targ&\ctrl{1}&\qw&\qw&\qw&\qw&\qw&\ctrl{1}&\targ&\qw&\targ&\ctrl{1}&\qw&\qw&\qw&\qw&\qw&\ctrl{1}&\targ&\qw&\qw\\
&\qw&\qw&\targ&\ctrl{1}&\qw&\qw&\qw&\ctrl{1}&\targ&\qw&\qw&\qw&\targ&\ctrl{1}&\qw&\qw&\qw&\ctrl{1}&\targ&\qw&\qw&\qw\\
&\qw&\qw&\qw&\targ&\ctrl{1}&\qw&\ctrl{1}&\targ&\qw&\qw&\qw&\qw&\qw&\targ&\ctrl{1}&\qw&\ctrl{1}&\targ&\qw&\qw&\qw&\qw\\
&\gate{H}&\qw&\qw&\qw&\targ&\gate{R_Z(\delta\frac{t}{2})}&\targ&\qw&\qw&\qw&\gate{HSH}&\qw&\qw&\qw&\targ&\gate{R_Z(\delta\frac{t}{2})}&\targ&\qw&\qw&\qw&\gate{HS^\dagger}&\qw
}
\end{equation}
\end{widetext}
where the first is the qubit corresponding to single orbital $p$ and the last one is $q$. This can be further reduced if arbitrary connectivity is allowed (see eg.~\cite{Hastings2015}). In total, one finds at most $24M$ rotations and less than $48M^2$ nearest neighbour CNOT gates. The same estimates also hold for $U_4$.
Using the fermionic-swap network algorithm instead, we can implement $U_3$ and $U_4$ by performing $2M(4M-1)$ two-qubit fermionic simulation gates with parallel depth $4M$, each one requiring at most $5$ rotations and $3$ CNOT (since we are not implementing evolution under the on site interaction at the same time, this reduces to just $2$ arbitrary rotations~\cite{Vatan2004}). Most of these are simple fermionic swap gates 
\begin{equation}
\mathrm{fSWAP} = \begin{pmatrix}
1&0&0&0\\
0&0&1&0\\
0&1&0&0\\
0&0&0&-1\\
\end{pmatrix}\;,
\end{equation}
requiring $3$ CNOT and additional Clifford gates. The number of (controlled) arbitrary rotations is therefore at most equal to the serial count for the naive implementation or $8M$ in the parallel case. The gate cost of implementing the propagators $U_3$ and $U_4$ is summarized in Tab.~\ref{tab:hop_u3}.

%\FIXME{TODO: in 3D it should be possible to perform the circuit with depth $\propto 4M^{\frac{2}{3}}$, but only when open boundary conditions are chosen, should we comment on it?}
\begin{table}[]
\begin{tabular}{|l|cc|cc|}
	\hline
 & \multicolumn{2}{c|}{\# c-Rz} & \multicolumn{2}{c|}{\# CNOT} \\
 &    serial       &    parallel       &    serial       &   parallel        \\ \hline
naive &     $<$24M      &  $<$24M      &   $<48M^2$        &   $<48M^2$       \\
fermionic swap &    $<$24M      &     8M      &   6M(4M-1)        &   12M       \\ \hline
\end{tabular}
\caption{Gate cost for the $U_3/U_4$ propagator. See also Fig.~\ref{fig:hop_gcount} for tighter estimates of the naive cost.}
\label{tab:hop_u3}
\end{table}

\begin{figure}
    \centering
    \includegraphics[scale=0.33]{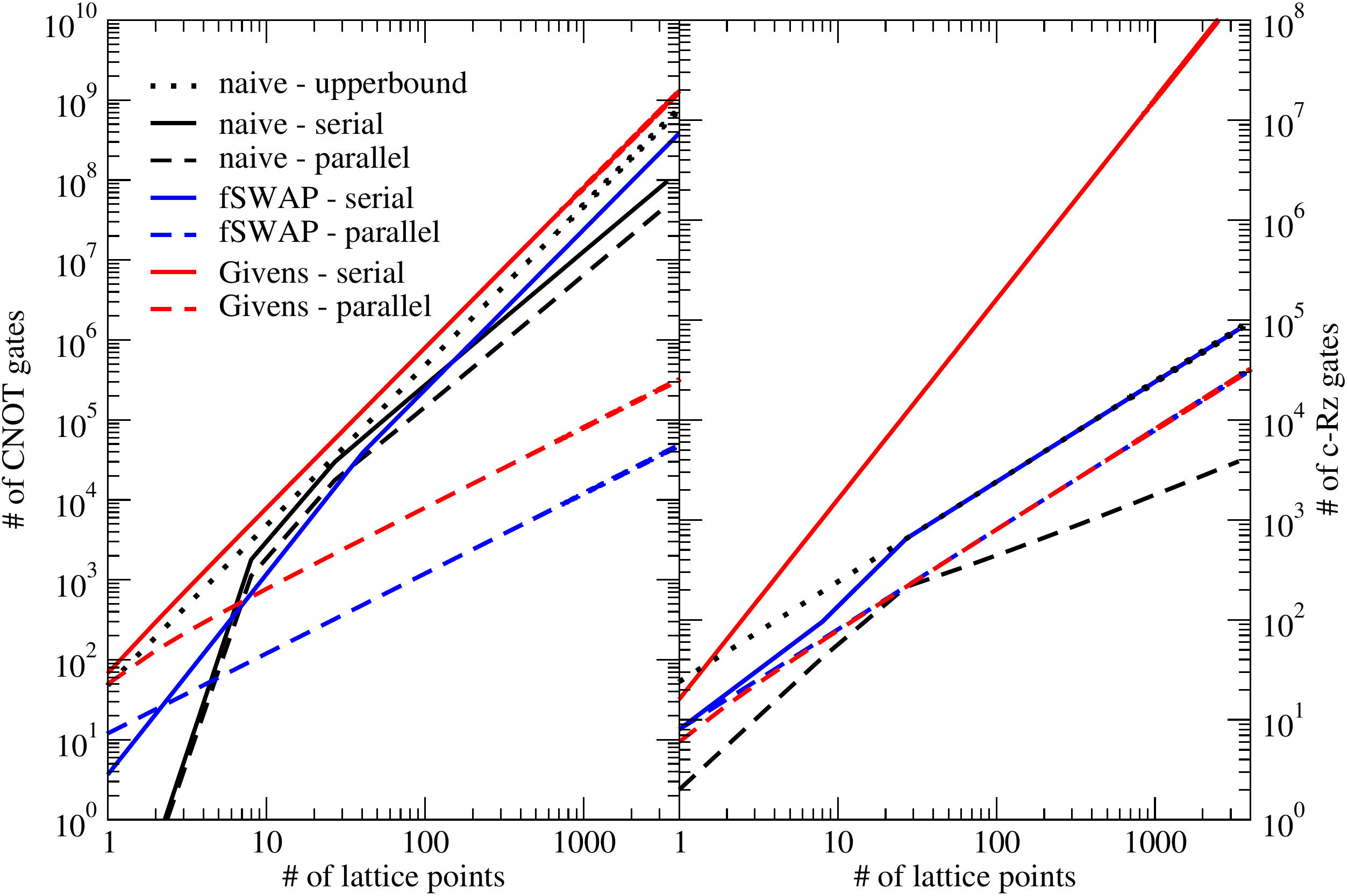}
    \caption{Empirical gate counts from a simulation of the naive implementation of $U_3/U_4$ for system sizes $M\in(8,3375)$. Also shown the gate cost estimates for the fSWAP algorithm and the implementation of $U_2$ using Givens rotations (cf.~Tab.~\ref{tab:hop_u3})}
    \label{fig:hop_gcount}
\end{figure}

In Fig.~\ref{fig:hop_gcount} we present the empirical gate counts for the evolution operator $U_3/U_4$ using both the naive implementation, the fermionic swap network and the exact implementation using Givens rotations (for the latter one we count half the cost of non-controlled rotations). We see that the latter two approaches provide a considerable reduction in CNOT counts but at the price of rising the parallel depth for the rotation gates, this might have an impact on the fault-tolerant implementation of the algorithm.
 
\section{Details on qubitization}
\label{app:qubitiz}
%\FIXME{Here I will just sketch the idea, if we want details I have notes I can use to add material.}
The basic idea behind qubitization~\cite{Low_and_Chuang_2019} is to represent the system Hamiltonian in the following way:
\begin{equation}
H = \lambda \left(\langle G\lvert\otimes\mathbb{1}\right) \mathrm{select}(H)\left(\mathbb{1}\otimes\rvert G\rangle\right)
\end{equation}
where the first register holds an ancilla space of dimension $>\Gamma$ where
\begin{equation}
H = \sum_{j=0}^{\Gamma-1} \lambda_j H_j\;,
\end{equation}
the coefficients $\lambda_j>0$ and $H_j$ are Clifford operations. For ease of derivation, we will assume that $H_0\equiv\mathbb{1}$ and if the original Hamiltonian was traceless, we add it while increasing $\Gamma$ accordingly.
The two main subroutines we need to perform qubitization are the {\it select} unitary $V_S$ and the {\it prepare} unitary $V_P$ which can be respectively defined as
\begin{equation}
\label{eq:select_def}
V_S = \sum_{j=0}^{\Gamma-1} \rvert j\rangle\langle j\lvert \otimes H_j
\end{equation}
and
\begin{equation}
\label{eq:prepare}
V_P\ket{0}\rvert G\rangle = \frac{1}{\sqrt{\lambda}}\sum_{j=0}^{\Gamma-1}\sqrt{\lambda_j}\rvert j\rangle \quad \lambda = \sum_{j=0}^{\Gamma-1} \lambda_j\;.
\end{equation}
As mentioned in the main text, the central object of this scheme is the qubiterate unitary defined in Eq.~\eqref{eq:qubiterate} of the main text and which can be implemented using the two basic unitaries defined above as 
\begin{equation}
\label{eq:qubiterate_2}
Q = -i \left(\left(2\rvert G\rangle\langle G\lvert-\mathbb{1}\right)\otimes\mathbb{1}\right) V_S
\end{equation}
whose eigenvalues are
\begin{equation}
\eta_{\pm} = \mp e^{\pm i arcsin\left(\eta\right)}
\end{equation}
where $H/\lambda=\sum_\eta \eta\rvert\eta\rangle\langle\eta\lvert$ is the spectral decomposition of the (scaled) hamiltonian. The eigenvectors of the qubiterate are
\begin{equation}
\rvert\eta_\pm\rangle = \frac{1}{\sqrt{2}}\left(\rvert G_\eta\rangle\pm i \rvert G^\perp_\eta\rangle\right)\,,
\end{equation}
which are connected with the energy eigenstate by
\begin{equation}
\rvert G_\eta\rangle = \rvert G\rangle\otimes\rvert \eta\rangle
\end{equation}
and
\begin{equation}
\rvert G^\perp_\eta\rangle = \frac{\eta\rvert G_\eta\rangle-\mathrm{select}(H)\rvert G_\eta\rangle }{\sqrt{1-\eta^2}}\;.
\end{equation}

\subsection{Gate cost of the qubiterate}
\label{app:qub_cost}
We now proceed to estimate the qubiterate gate cost. 
Following the discussion in the main text in order to implement the sequence of controlled qubiterates, we need to implement an initial controlled prepare (whose small cost we neglect in the estimates that follow) and then for every qubiterate we need one {\it select} and two {\it prepare} without controls and one controlled reflection. Since we are trying to provide a lower bound on the gate count and techniques that only need one copy of {\it prepare} per step are known~\cite{Babbush2018}, we will only count the cost of one prepare per step.

For our model with $N_f=4$ fermionic species, the kinetic energy requires $\Gamma_K=24M$ and the potential energy part needs $\Gamma_V=14M$ for a total $\Gamma=38M$. The size of the ancilla register required to encode the flag state $\ket{G}$ is thus $N_A=\lceil \log_2(\Gamma)\rceil$. The easiest unitary to implement is the {\it prepare} operation $V_P$ defined by
\begin{equation}
V_P\left(\ket{0}\right) = \ket{G} =\frac{1}{\sqrt{\lambda}}\sum_{j=0}^{\Gamma-1}\sqrt{\lambda_j}\rvert j\rangle \;.
\end{equation}
Without assuming any structure in the coefficients $\{\lambda_j\}$, we can always prepare the flag state $\ket{G}$ on the ancilla register with $N_A$ qubits using at most (see eg.~\cite{Bergholm2005})
\begin{itemize}
    \item $2^{N_A}-2N_A-2$ CNOT
    \item $2^{N_A}-N_A-2$ one qubit gates
\end{itemize}
and to be conservative, we will count $3$ z-rotations per one-qubit unitary. The second unitary we need is the controlled reflection ${}_c\Pi_0$ acting on the $N_A$ ancillas plus the control which we implement as in~\cite{Childs2018} using a multiply controlled Z gate with $N_A$ controls implemented using the ancilla-based scheme described in~\cite{Maslov2016} which needs
\begin{itemize}
    \item $\lceil\frac{N_A-2}{2}\rceil$ ancillas in $\ket{0}$
    \item $8N_A-9$ T gates
    \item $6N_A-6$ CNOT gates
    \item $4N_A-6$ Hadamard gates
\end{itemize}

Finally we can implement the control circuit for the {\it select} operation $V_S$ using the optimized scheme from \cite{Childs2018} which needs $N_B=N_A-1$ additional ancilla qubits prepared and returned in $\ket{0}$ and 
\begin{itemize}
    \item 2 NOT gates
    \item $2^{N_A+1}+2N_A-8$ Hadamard gates
    \item $2^{N_A-1}-N_A$ Phase gates
    \item $15 2^{N_A-1}+6N_A-28$ $T/T^\dagger$ gates
    \item $15 2^{N_A-1}+6N_A-26$ CNOT gates
\end{itemize}
Note, however, that this control circuit will cycle over all the possible $2^{N_A}$ possible states of the control register and we can terminate this only after the needed $\Gamma$ are obtained. In order to estimate this uncertainty of the analytical gate count, we will make use of the relation $2^{N_A}\geq\Gamma>2^{N_A-1}$ to bound the gates cost.

On top of this, we need to implement all the $\Gamma$ controlled unitaries that are, however, all Clifford operations.
Here, we neglect them to estimate a lower bound of the gate cost, which is sufficient for the comparison between qubitization and Trotter decompositions.
% As motivated in the main text, we neglect them in the cost estimate since it is already too large.
% \FIXME{rewrite this sentence better}
Counting only two-qubit Clifford gates and rotations or T gates, and reusing ancillas for both {\it select} and the reflection, we find the following cost estimate for a single application of the controlled qubiterate
\begin{itemize}
    \item $2N_A - 1$ ancilla qubits
    \item $17 2^{N_A-1}+10N_A-34$ CNOT gates
    \item $2^{N_A}-N_A-2$ $U(2)$ gates
    \item $15 2^{N_A-1}+14N_A-37$ $T/T^\dagger$ gates
\end{itemize}
The gate cost of qubitization is compared with that of Trotter decompositions in Fig.~\ref{fig:gcount_beta} in the main text.
Note that $T$ gates can be implemented using arbitrary Z-rotations, and hence we count $T$ gates as $R_Z$ gates for the estimates there.
%\FIXME{Justify better}.

% \FIXME{We might want to do something with these....}
% \ale{Assuming $10$ ns gates and no overhead from error correction the estimates above translate into the red line shown in Fig.~\ref{fig:timings}. Explain what the other curves are and that we can get an $x100$ speedup using qubitization.}
% \andy{We should just refer to Fig.~\ref{fig:total_cost_woth_qub} since all messages are indeed conveyed there.}

% \begin{figure}
%     \centering
%     \includegraphics[scale=0.33]{images/timingsD-eps-converted-to}
%     \caption{Estimated wall clock time for execution of the phase estimation kernel of the linear response algorithm of \cite{Roggero2018} as a function of nucleon number. The different curves correspond to different implementations of the time evolution operator.\ale{Double check results and remove "preliminary".}}
%     \label{fig:timings}
% \end{figure}

% \begin{figure}
%     \centering
%     \includegraphics[scale=0.3]{images/tot_time_beta_with_qub-eps-converted-to}
%     \caption{New version of Fig.~\ref{fig:timings}. \FIXME{Done in a sloppy way by estimating $15$ T gates per rotation.}}
% \end{figure}

\end{document}